\documentclass[a4paper,12pt,dvipsnames,usernames]{article}%
\pdfoutput=1

\usepackage{jheppub}
\usepackage{booktabs}
\usepackage{graphicx} 
\usepackage{amsmath} 
\usepackage{amssymb}
\usepackage{bm} 
\usepackage{paralist,array} 
\usepackage{units}
\graphicspath{{figures/}} \usepackage{slashed} \usepackage{todonotes}
\usepackage[parfill]{parskip} \usepackage{longtable}
\usepackage{xspace}
\usepackage{bbm}
\newcommand{\BR}{\mathop{\mathrm{BR}}}
\newcommand{\BRinv}{\text{BR}_\text{inv}}
\newcommand{\BRfid}{\text{BR}_\text{fid}}
\newcommand{\decay}{\text{decay}}
\newcommand{\geom}{\text{geom}} 
\newcommand{\faser}{\textsc{faser}}
\newcommand{\FASER}{FASER\xspace}
\renewcommand{\det}{\text{det}}

\title{Light scalar production from Higgs bosons and \FASER2}

\author[1]{Iryna Boiarska,}
\author[2]{Kyrylo~Bondarenko,}
\author[2]{Alexey~Boyarsky,}
\author[2]{Maksym~Ovchynnikov,}
\author[1]{Oleg~Ruchayskiy,}
\author[3]{Anastasia~Sokolenko}

\emailAdd{boiarska@nbi.ku.dk}
\emailAdd{bondarenko@lorentz.leidenuniv.nl}
\emailAdd{boyarsky@lorentz.leidenuniv.nl}
\emailAdd{ovchynnikov@lorentz.leidenuniv.nl}
\emailAdd{oleg.ruchayskiy@nbi.ku.dk}
\emailAdd{anastasia.sokolenko@fys.uio.no}

\affiliation[1]{Discovery Center, Niels Bohr Institute, Copenhagen
University, Blegdamsvej 17, DK-2100 Copenhagen, Denmark}
\affiliation[2]{Intituut-Lorentz, Leiden University, Niels Bohrweg 2, 2333
CA Leiden, The Netherlands}
\affiliation[3]{Department of Physics, University of Oslo, Box 1048, NO-0371, Oslo, Norway}

\begin{document}

\abstract{The most general renormalizable interaction between the Higgs sector and a new gauge-singlet scalar $S$ is governed by two interaction terms: cubic and quartic. The quartic term is only loosely constrained by invisible Higgs decays and given current experimental limits about $10\%$ of all Higgs bosons at the LHC can be converted to new scalars with masses up to $m_{\rm Higgs}/2$. By including this production channel, one significantly extends the reach of the LHC-based Intensity Frontier experiments. We analyze the sensitivity of the FASER experiment to this model and discuss modest changes in the \FASER2 design that would allow exploring an order-of-magnitude wider part of the Higgs portal's parameter space.}
\maketitle

\section{Introduction: scalar portal and FASER experiment}
The Standard Model of particle physics (SM) is extremely successful in
explaining accelerator data. Yet it fails to explain several observed
phenomena: neutrino masses, dark matter and baryon asymmetry of the Universe. To explain these phenomena, we need to postulate new particles that should not nevertheless spoil extremely successful Standard Model predictions. These new hypothetical particles can be heavy, thus evading detection at $\sqrt s = 13$~TeV collision energy of the LHC. Such particles would induce higher-dimensional (non-renormalizable) interactions with SM fields, the signatures of such operators are being searched at the LHC (see \textit{e.g.}~\cite{Brivio:2017vri} for a review).

Alternatively, new particles can be light yet have very weak couplings to the Standard Model -- feebly interacting particles, or FIPs.
In this case, their interaction with the SM can be governed even by relevant (dimensions 3 and 4) operators with small couplings.
Such models are generically called \textit{portals} because  trough such operators FIPs  can mediate interactions with some ``\textit{dark sectors}'' -- other new particles that otherwise are inaccessible. 

In this paper, we consider the most general form of the scalar (or Higgs) portal~\cite{McDonald:1993ex,Burgess:2000yq,Patt:2006fw,OConnell:2006rsp} that has been the subject of active analysis in the recent years, see \emph{e.g.}~\cite{Djouadi:2012zc,Curtin:2013fra,Alekhin:2015byh,Arcadi:2019lka} and refs. therein.  Namely, we introduce a scalar
particle $S$ that carries no Standard Model charges and interacts with the Higgs doublet $H$ via
\begin{equation}
  \mathcal{L} = \mathcal{L}_{SM} + \frac{1}{2} (\partial_{\mu} S)^2 + 
  (\alpha_1 S + \alpha_{2} S^2) \left(H^{\dagger} H -\frac{v^2}{2}\right) -\frac{m_S^2}{2}S^2,
  \label{eq:L1}
\end{equation}
where $v$ is the Higgs VEV and the model is parametrized by three new
constants: $\alpha_1, \alpha_2$ and the scalar mass $m_S$.  
After electroweak symmetry breaking, the $SHH$ interaction~\eqref{eq:L1} leads to a quadratic mixing between $S$ and the Higgs boson $h$. Transforming the Higgs field into the mass basis,
$h \to h + \theta S$ ($\theta \ll 1$), one arrives at the following Lagrangian, describing interactions of the new boson $S$ with the
SM fermions, intermediate vector bosons and the Higgs boson:
\begin{equation}
  \mathcal{L}_{SM}^{S} = -\theta\frac{m_f}{v} S \bar{f}f +
    2\theta\frac{ m_W^2}{v} S W^+ W^- +
    \theta\frac{m_Z^2}{v} S Z^2 + \frac{\alpha}{2}S^2 h + \dots
    \label{g01}
\end{equation}
where $\dots$ denote quartic and higher terms. The interactions~\eqref{g01} also mediate effective couplings of the scalar to photons, gluons, and flavor changing quark operators~\cite{Boiarska:2019jym}, opening many production
channels at both LHC and Intensity Frontier experiments. The phenomenology of light GeV-like scalars has been worked out in~\cite{Bird:2004ts,Batell:2009jf,Bezrukov:2009yw,Clarke:2013aya,Schmidt-Hoberg:2013hba,Evans:2017lvd,Bezrukov:2018yvd,Monin:2018lee,Winkler:2018qyg,Frugiuele:2018coc,Helmboldt:2016zns} as well as in~\cite{Voloshin:1985tc,Raby:1988qf,Truong:1989my,Donoghue:1990xh,Willey:1982ti,Willey:1986mj,Grzadkowski:1983yp,Leutwyler:1989xj,Haber:1987ua,Chivukula:1988gp} in the context of the light Higgs boson.
Most of these works concentrated on the Lagrangian with  $\alpha_1 = 0$
in which case the couplings $\theta$ and $\alpha$ in~\eqref{g01} become
related.\footnote{Alternative class of models has super-renormalizable interaction only between the Higgs boson and the scalar ($\alpha_2 = 0$), see~\cite{Fradette:2018hhl} and refs.\ therein. In this case, of course, there is no $S^2 h$ term in the Lagrangian~\eqref{g01}.}
In this work we consider $\alpha_1 \neq 0$.  Phenomenologically, this allows to decouple \emph{decay channels} (controlled by $\theta$) and \emph{production channels} (controlled by $\alpha$), \textit{c.f.}~\cite{Fradette:2017sdd} where phenomenology of such a model is also discussed.
As we will see below, the parameter $\alpha$ is only weakly constrained by the invisible Higgs decays~\cite{Sirunyan:2018owy,Aaboud:2018sfi} and can be quite sizeable (if unrelated to $\theta$).
As a result, the production via $h \to SS$ process becomes possible and is operational for scalar masses up to $m_h/2$ which allows to significantly extend the sensitivity reach of the LHC-based experiments.

We note that the production channel via the off-shell Higgs
bosons (\textit{e.g.}\ coming from neutral meson decays, such as 
$B_s \to SS$ for $2m_S < m_B$) starts to dominate over production via flavour
changing mixing for $\theta^2 < 10^{-9} \div 10^{-10}$, see~\cite{Boiarska:2019jym}.
We will not consider this effect in the current work, mostly concentrating on $m_S \gtrsim \unit[5]{GeV}$.

Searches for light scalars have been previously performed by
CHARM~\cite{Bergsma:1985qz}, KTeV~\cite{AlaviHarati:2000hs},
E949~\cite{Artamonov:2008qb,Artamonov:2009sz}, 
Belle~\cite{TheBelle:2015mwa,Seong:2018gut},
BaBar~\cite{Lees:2013kla},
LHCb~\cite{Aaij:2015tna,Aaij:2016qsm},
CMS~\cite{Sirunyan:2018owy, Sirunyan:2018mot,Sirunyan:2018mgs} and 
ATLAS~\cite{Aad:2015txa,Aaboud:2018sfi, Aaboud:2018esj,Aaboud:2018iil}
experiments. Significant progress in searching for light scalars can be
achieved by the proposed and planned intensity-frontier experiments such as
SHiP~\cite{Alekhin:2015byh,Anelli:2015pba,Bondarenko:2019yob},
CODEX-b~\cite{Gligorov:2017nwh},
MATHUSLA~\cite{Chou:2016lxi,Evans:2017lvd,Curtin:2018mvb,Bondarenko:2019yob}, FASER~\cite{Feng:2017uoz,Feng:2017vli}, SeaQuest~\cite{Berlin:2018pwi},
NA62~\cite{Mermod:2017ceo,CortinaGil:2017mqf,Drewes:2018gkc} and a number of other experiments (see~\cite{Beacham:2019nyx} for an overview). The summary of the current experimental status of the light scalar searches is provided in the Physics Beyond Collider report~\cite{Beacham:2019nyx}.

\begin{figure}[t!]
  \centering
  \begin{minipage}[b]{0.48\textwidth}
    \includegraphics[width=\textwidth]{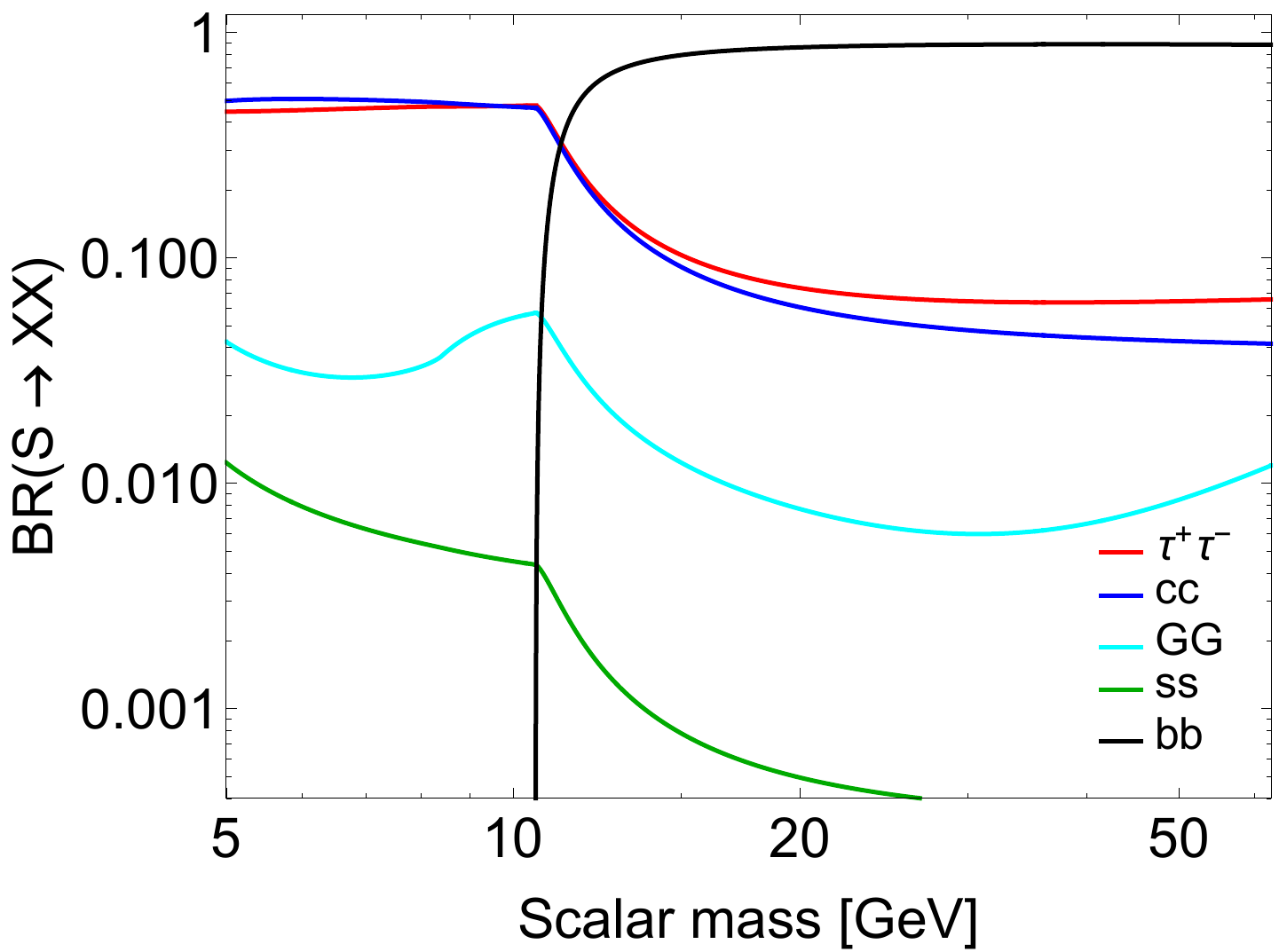}
  \end{minipage}
  \hfill
  \begin{minipage}[b]{0.48\textwidth}
    \includegraphics[width=\textwidth]{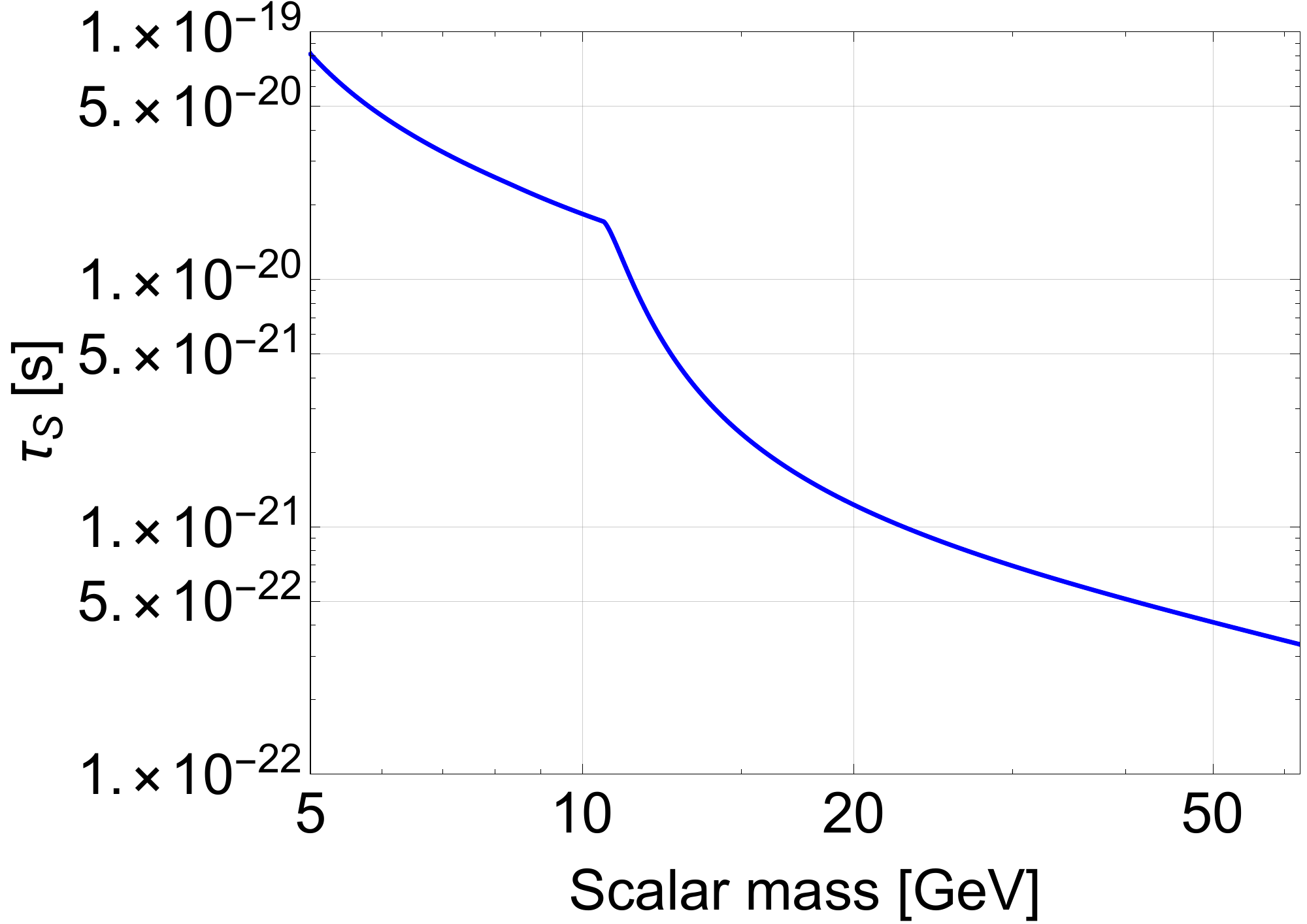}
  \end{minipage}
  \caption{\emph{Left panel:} branching ratios of the decays of a scalar $S$ as a function of its mass.
    We use perturbative decays into quarks and gluons (see~\cite{Boiarska:2019jym} for details).
    \emph{Right panel:} the lifetime of a scalar $S$ as a function of its mass for the mixing angle $\theta^2 = 1$.
    The lifetime is obtained using decays into quarks and gluons (and $\tau$'s) within the framework of perturbative QCD.}
  \label{fig:decays}
\end{figure}

\subsection{Existing bounds}
\label{sec:bounds}

The up to date experimental constraints in the $m_S$-$\theta$ plane can be found in the scalar portal section of~\cite{Beacham:2019nyx}.
The strongest experimental constraints on the parameter $\alpha$ come from the invisible Higgs decay. In the Standard Model the decay $h \to ZZ \to 4\nu$ has the branching ratio $\mathcal{O}(10^{-3})$.
Current limits on the Higgs to invisible are $ \BRinv < 0.19$ at $95\%$ CL~\cite{Sirunyan:2018owy}. Future searches at LHC Run~3 and at the High-Luminosity (HL) LHC (HL-LHC, Run 4) are projected to have sensitivity at the level $\BRinv \sim 0.05 \text{ --- } 0.15$ at $95\%$~CL \cite{Bechtle:2014ewa} maybe going all the way to a few
\textit{percents}~\cite{deBlas:2019rxi}. In what follows we will assume that the branching ratio $\BRinv$ is saturated by the $h\to SS$ decay.
Using
\begin{equation}
  \label{eq:3}
  \Gamma_{h\to SS} = \frac{\alpha^2}{32\pi m_h} \sqrt{1-\frac{4m_S^2}{m_h^2}}
\end{equation}
we obtain the corresponding value of $\alpha^2 \sim  \unit[5]{GeV^2}$ for $m_{S} \ll m_{h}$.

Apart from the invisible Higgs decays, the ATLAS and CMS collaborations have previously performed studies of the $h \to SS \to 4b$, $h \to SS \to 2b2\mu$, $h \to SS \to 2\tau 2\mu$, $h \to SS \to 2\tau 2b$,
etc.~\cite{Aaboud:2018esj,Aaboud:2018iil,Sirunyan:2018mot,Sirunyan:2018mgs,Sirunyan:2018pzn,Sirunyan:2018mbx} for the light (pseudo)scalar in the mass ranging between $\mathcal{O}(10)$~GeV and $m_h/2$. The obtained constraints, however, do not restrict the parameters relevant for the \FASER2 experiment as they search for prompt decays of the scalars, while in our model the $c\tau_S \sim \mathcal{O}(100)$~meters.

\subsection{The FASER experiment}
\label{sec:faser}

\begin{figure}[t]
  \centering \includegraphics[width=0.95\textwidth]{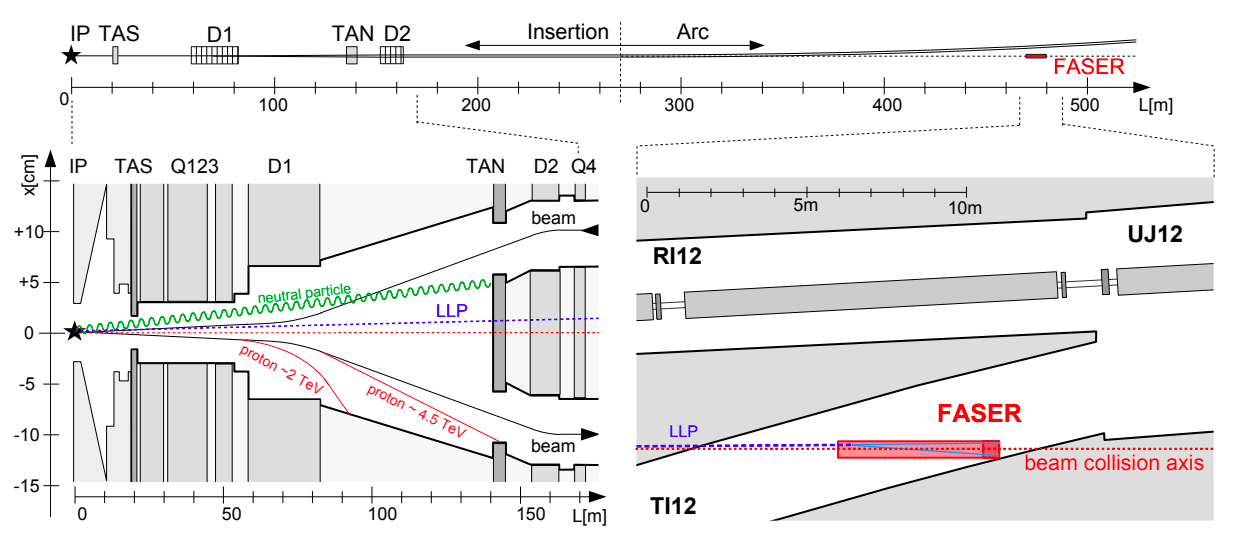}
  \caption{The scheme of the FASER experiment. The figure from~\cite{Ariga:2019ufm}.}
  \label{fig:faser_exp}
\end{figure}

FASER (\textit{ForwArd Search ExpeRiment}, Fig.~\ref{fig:faser_exp}), is an Intensity Frontier experiment dedicated to searching for light, extremely weakly-interacting particles that may be produced in the LHC's high-energy collisions in the far-forward region and then travel long distances without interacting~\cite{Feng:2017uoz,Ariga:2018pin,Ariga:2018uku,Ariga:2018zuc,Ariga:2019ufm}.
\FASER is approved to collect data in 2021-2023 during the LHC Run 3.
If \FASER is successful, \FASER2, a much larger successor, could be constructed in Long Shutdown 3 and collect data during the High-Luminosity  Run 4 in 2026-2035.
The relevant parameters of \FASER and \FASER2 are shown in Table~\ref{tab:FASER-parameters}.
We also list the alternative configuration of \FASER 2 which we will use for comparison in this work.

\begin{table}[t!]
  \centering
  \begin{tabular}{p{12em}|c|c|c|c|c}
    \toprule
    \bf Phase & $\mathcal{L}$, $\text{fb}^{-1}$ & $L$ [m] & $R$ [m] & $l_{\text{det}}$ [m] & $\theta_{\faser}$ [rad] \\ 
    \toprule
    \FASER & 150 & 480 & 0.1 & 1.5 & $2.1 \cdot 10^{-4}$ \\ \midrule
    \FASER2 & 3000 & 480 & 1 & 5 & $2.1 \cdot 10^{-3}$ \\ \midrule
    \FASER2\newline (alternative  configuration) & 3000 & 480 & 1.5 & 5 & $3.1 \cdot 10^{-3}$ \\ 
    \bottomrule
  \end{tabular}
  \caption{Parameters of the \FASER experiment.
  Prototype detector (\FASER) is approved to collect data during the LHC Run 3. \FASER2 is planned for HL-LHC phase, but its configuration is not finalized yet. In the third line, we propose an alternative configuration of \FASER2 that would allow drastically increasing its reach towards the scalar portal. $\mathcal{L}$ is the integrated luminosity of the corresponding LHC run. $L$ is the distance between the ATLAS interaction point and the entrance of the \FASER decay vessel. $R$ is the radius of the decay vessel. $l_{\text{det}}$ is the length of the  detector and $\theta_{\text{FASER}} = R/L$ is the angle, so that the solid angle subtended by the detector is given by $\Omega_\faser = \pi \theta_\faser^2$. For our investigation, we assume that the decay vessel is a cylinder, centered around the beam axis.}
  \label{tab:FASER-parameters}
\end{table}

While the design of the first phase is fixed, the \FASER2 is not finalized yet. \emph{We demonstrate therefore how the parameters of the future \FASER2 experiment will affect its sensitivity.}

The paper is organized as follows:
\begin{compactitem}[--]
  \item In Section~\ref{sec:scalars} we estimate the number of decay events in the \FASER\ detectors. This Section allows for easy cross-check of our main results and gives the feeling of the main factors that affect the sensitivity. 
  \item In Section~\ref{sec:results} we outline our estimates based on which the conclusion is drawn. We also demonstrate that an increase of the geometric acceptance by the factor $\sim 2$ (\textit{e.g.} via increase of the radius of the decay vessel of  \FASER2 from 1~m to 1.5~m) would allow a wide region of the parameter space to be probed.
  \item Appendices provide some details of our computations that would permit the interested reader to reproduce them.
\end{compactitem}

\section{Scalars from Higgs bosons}
\label{sec:scalars}

\subsection{Naive estimate: what can be expected?}
\label{sec:production}

Before running MC simulations (and to have a way to verify the
simulation results) we start with analytic estimates of the sensitivity of \FASER2. The number of detected events is given by the following formula~\cite{Bondarenko:2019yob}:
\begin{equation}
  \label{eq:1}
  N_\det = N_S \times \epsilon_\geom \times P_\decay \times \epsilon_\det.
\end{equation}
Here, $N_{S}$ is the number of scalars produced at the LHC experiment; in our case $N_S = 2 N_h \BR(h\to SS)$, $N_h$ -- the number of produced Higgs bosons, $\epsilon_\geom$ is the \emph{geometric acceptance} -- the fraction of scalars whose trajectories intersect the decay volume, so that they \textit{could} decay inside it. The decay probability is given by the well-known formula
\begin{equation}
  P_{\decay}(l_{\decay}) = e^{-L/l_{\text{decay}}} - e^{-(L+l_{\det})/l_{\text{decay}}},
  \label{eq:Pdecay}
\end{equation}
where $L$ is the distance from the interaction point to the entrance of the fiducial volume, $l_{\det}$ is the detector length, and
$l_\decay = c\tau_S \beta_{S}\gamma_S$ is the decay length. Finally,
$\epsilon_{\det}\le 1$ is the \emph{detection efficiency} -- a fraction of all decays inside the decay volume for which the decay products could be detected. In the absence of detector simulations, we optimistically assume detector efficiency of \FASER\ to be $\epsilon_{\det}=1$.

The high luminosity LHC phase is expected to deliver $1.7\cdot 10^8$ Higgs bosons (the Higgs boson production cross-section at $\sqrt{s} = 13$~TeV is $\sigma_h \approx 55$~pb~\cite{Cepeda:2019klc}, going to $\unit[60]{pb}$ at 14~TeV). Further, we assume the fiducial Higgs decay to scalars equal to the lower bound of HL-LHC reach~\cite{Bechtle:2014ewa}:
\begin{equation}
    \label{eq:Brinv}
    \BRfid(h\to SS)  = 0.05.
\end{equation}

For the initial estimate of the number of produced scalars, we consider these Higgs bosons decaying at rest. In this case, we estimate the number of scalars flying into the solid angle of \FASER2 as
\begin{equation}
  \label{eq:N_naive}
  \epsilon_\geom^{\rm naive} = \frac{\Omega_\faser}{4\pi} \approx 1.1\times 10^{-6},
\end{equation}
where $\Omega_{\faser} = \pi\theta_{\faser}^{2}$, see Table~\ref{tab:FASER-parameters}. Plugging in the numbers we get $N_S^{\rm naive}= 2 N_h \times\epsilon_\geom^{\rm naive}\times \BRinv \approx 33$ scalars.  As most of the Higgs bosons fly along the beam axis, Eq.~\eqref{eq:N_naive} is a
strong \textit{underestimate} and we should expect a lot of scalars
\emph{flying through} the FASER fiducial volume.

For $l_{\det}\ll L$ (as it is the case for \FASER/\FASER2) the probability of decay~\eqref{eq:Pdecay} reaches its maximum for $l_{\text{decay}} \approx L$. The maximum is purely geometric, not related to the parameters of the scalar $S$ and numerically it is equal to
\begin{equation}
  P_{\text{decay}}^{(\max)}\simeq \frac{l_\det}{L} e^{-1} \approx 3.8\cdot 10^{-3},
  \label{eq:Pdecaymax}
\end{equation}
see also Fig.~\ref{fig:Higgs-distributions}.\footnote{The independence of the value~\eqref{eq:Pdecaymax} of the mass $m_{S}$ can be understood in the following way. Since the production of the scalar is independent on the coupling $\theta^{2}$ while the decay length depends on $\theta^2$, we can always adjust it for a fixed mass $m_{S}$ in a way such that $l_{\decay}(m_{S}, \theta^{2}) = L$. As we demonstrate below the values of $\theta^{2}$ for masses of interest (from few GeV to $m_h/2$) correspond to the region of the scalar parameter space that is currently unprobed by existing experiments.} Multiplying Eqs.~\eqref{eq:N_naive} and \eqref{eq:Pdecaymax} we find $\mathcal{O}(0.1)$ detectable events.
Given that this was a (strong) underestimate -- we see that more careful analysis is needed.
It will proceed as follows:
\begin{compactenum}[\bf 1.]
  
\item We start by assuming that all Higgs bosons travel along the beam axis, which allows for a much simplified analytic treatment.
  Then we comment on the effect of $p_T$ distribution of the Higgs bosons.
  
\item We determine the realistic geometrical acceptance $\epsilon_\geom \gg \epsilon_\geom^{\rm naive}$, since the actual angular distribution of scalars is peaked in the direction of the \FASER\ detector.

\item Finally, as scalars have non-trivial distribution in energy, for most of the scalars the decay probability is not equal to the maximal value, thus determining the width of the sensitivity area in the $\theta$ direction for a given mass.
\end{compactenum}

\subsection{Geometrical acceptance}
\label{sec:kappa}

\begin{figure}[t]
  \centering 
  \includegraphics[width=0.5\textwidth]{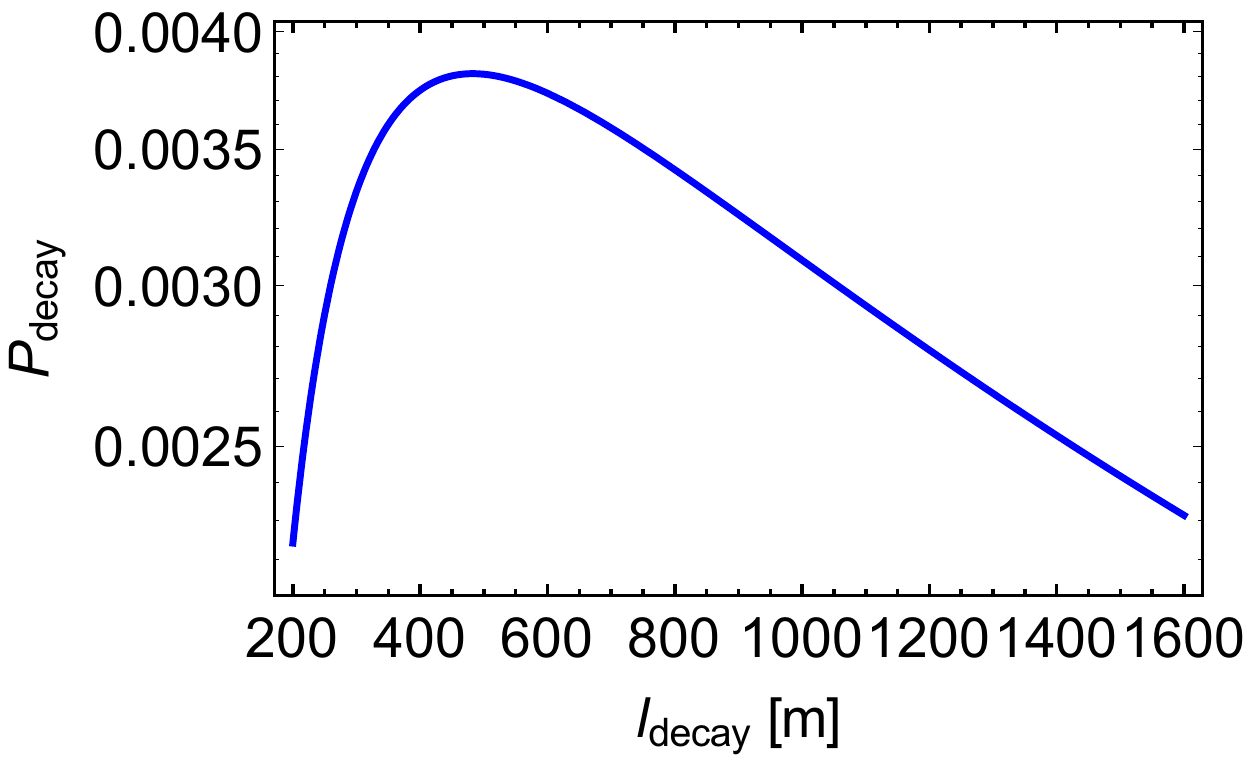}~\includegraphics[width=0.5\textwidth]{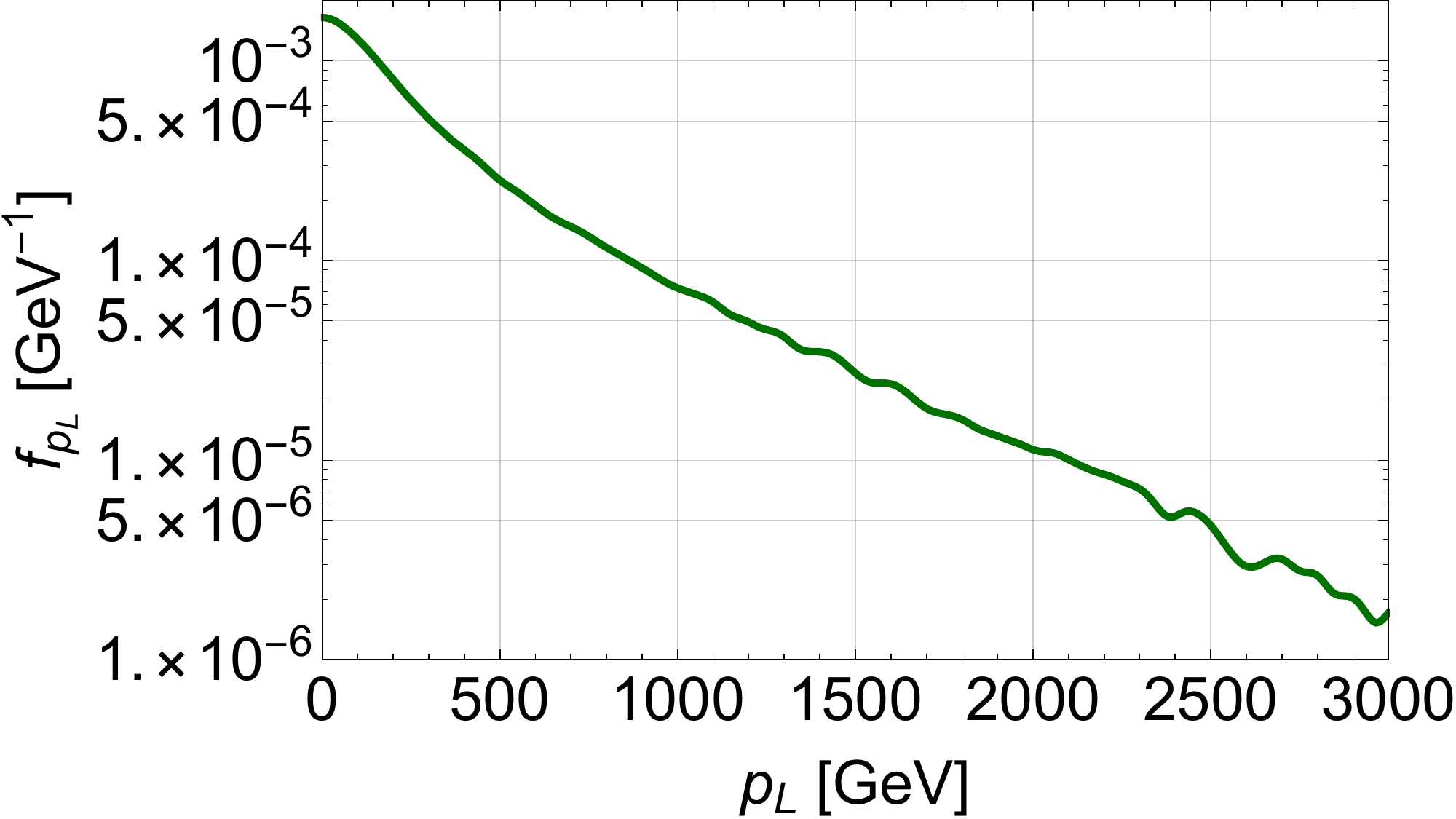}
  \caption{\textit{Left panel}: A probability of the scalar decay for \FASER2 as a function of a scalar's decay length $l_{\decay}$. \textit{Right panel}: the distribution function $f_{p_{L}} = \frac{1}{N_{h}}\frac{dN_{h}}{dp_{L}}$ of Higgs bosons by longitudinal momentum $p_L$. The simulations are based on MadGraph5\_aMC\@NLO~\protect\cite{Alwall:2014hca} and following~\protect\cite{Hirschi:2015iia}. See Appendix~\ref{sec:higgs} for details.}
  \label{fig:Higgs-distributions}
\end{figure}
Most Higgs bosons are traveling along the beam axis and therefore have $p_T \ll p_L$ (see Appendix~\ref{sec:higgs}).
Therefore, we perform the analytic estimates based on the purely longitudinal distribution of the Higgs bosons $f_{p_{L}} \equiv \frac{1}{N_{h}}\frac{dN_{h}}{dp_{L}}$ shown in Fig.~\ref{fig:Higgs-distributions}.

The angle $\theta_S$ between the scalar and Higgs boson directions in the laboratory frame is related to the scalar direction in the Higgs rest frame via
\begin{equation}
\hspace{-0.4cm}
  \tan\theta_S = \frac1{\gamma_h}\frac{\beta'_S \sin \theta'_S}{\beta'_S \cos\theta'_S + \beta_h}, 
  \label{eq:theta-thetaprime-relation}
\end{equation}
where
\begin{equation}
  \beta'_S = \sqrt{1 - \frac{4m_S^2}{m_h^2}}
  \label{eq:beta_S}
\end{equation}
is the velocity of a scalar in the rest frame of the Higgs boson, $\gamma_h$ and $\beta_h$ are Higgs boson's gamma factor and the velocity in the laboratory frame.\footnote{Although two scalars originate from each Higgs decay,  the  angle between the scalars in the laboratory frame is larger than $\theta_{\faser}$ unless $m_S$ is very close to $\frac{m_h}2$. In Appendix~\ref{sec:scalar-distr} we provide detailed estimates.}

\begin{figure}[t]
  \centering
    \includegraphics[width=0.5\textwidth]{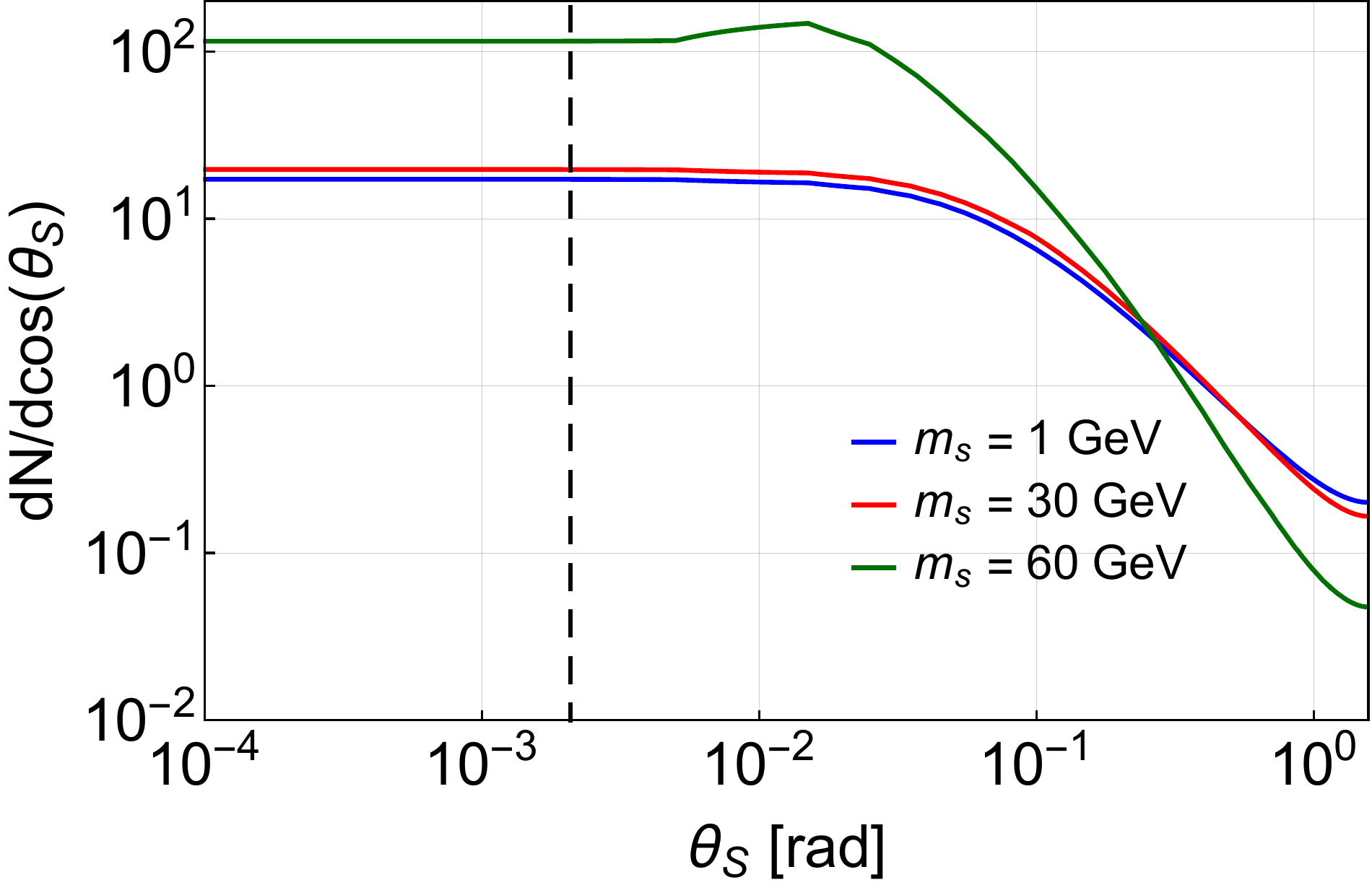}
  \caption{
The angular distribution of scalars for different scalar masses. The distribution is symmetric with respect to $\pi/2$ (right vertical axis). The vertical dashed line corresponds to $\theta_{S} = \theta_{\faser 2}$. The estimate is made under the assumption that Higgs bosons fly along the beam axis (see text for details).}
  \label{fig:thetaS_av}
\end{figure}

\begin{figure}[t]
  \centering
  \includegraphics[width=0.47\textwidth]{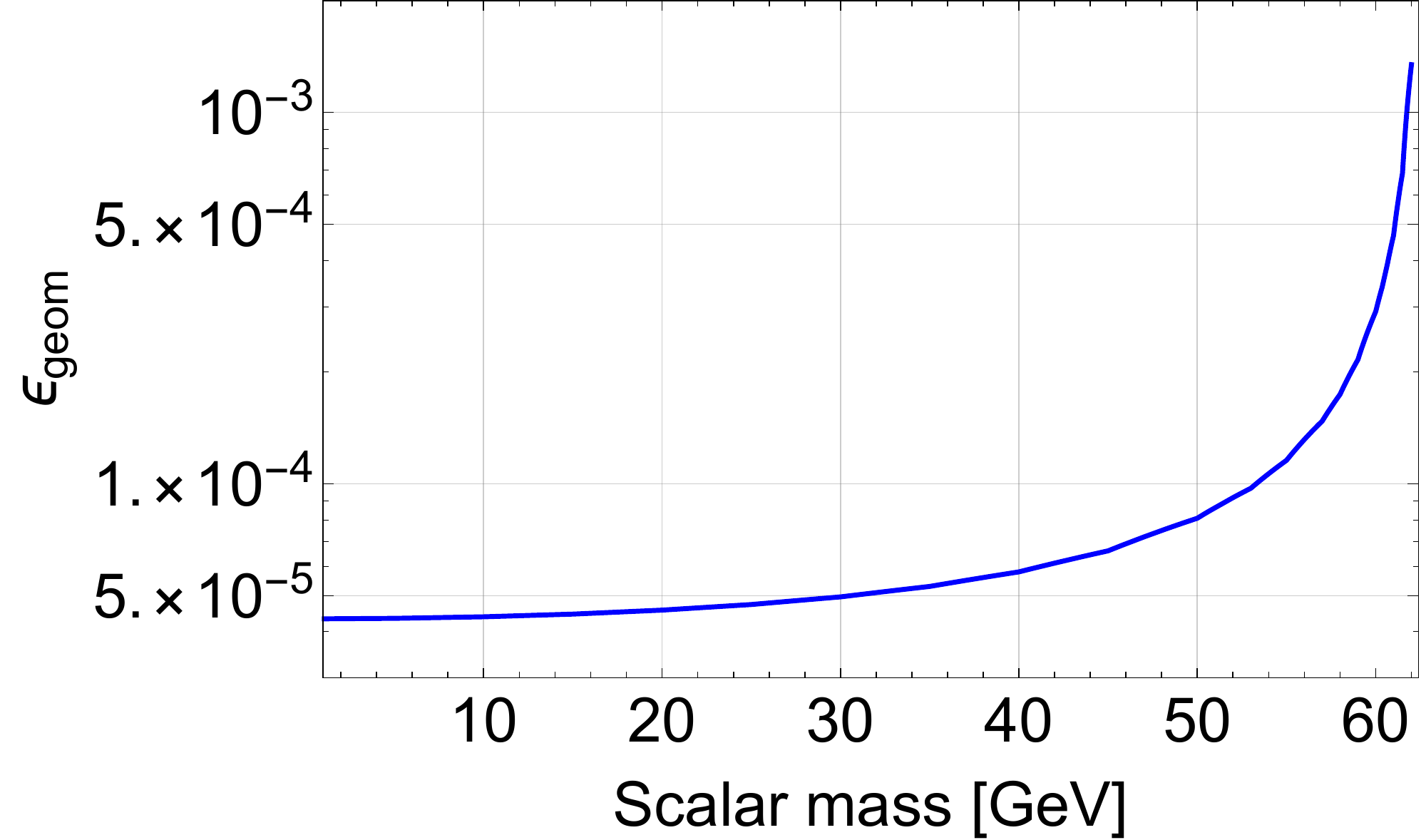}~\includegraphics[width=0.50\textwidth]{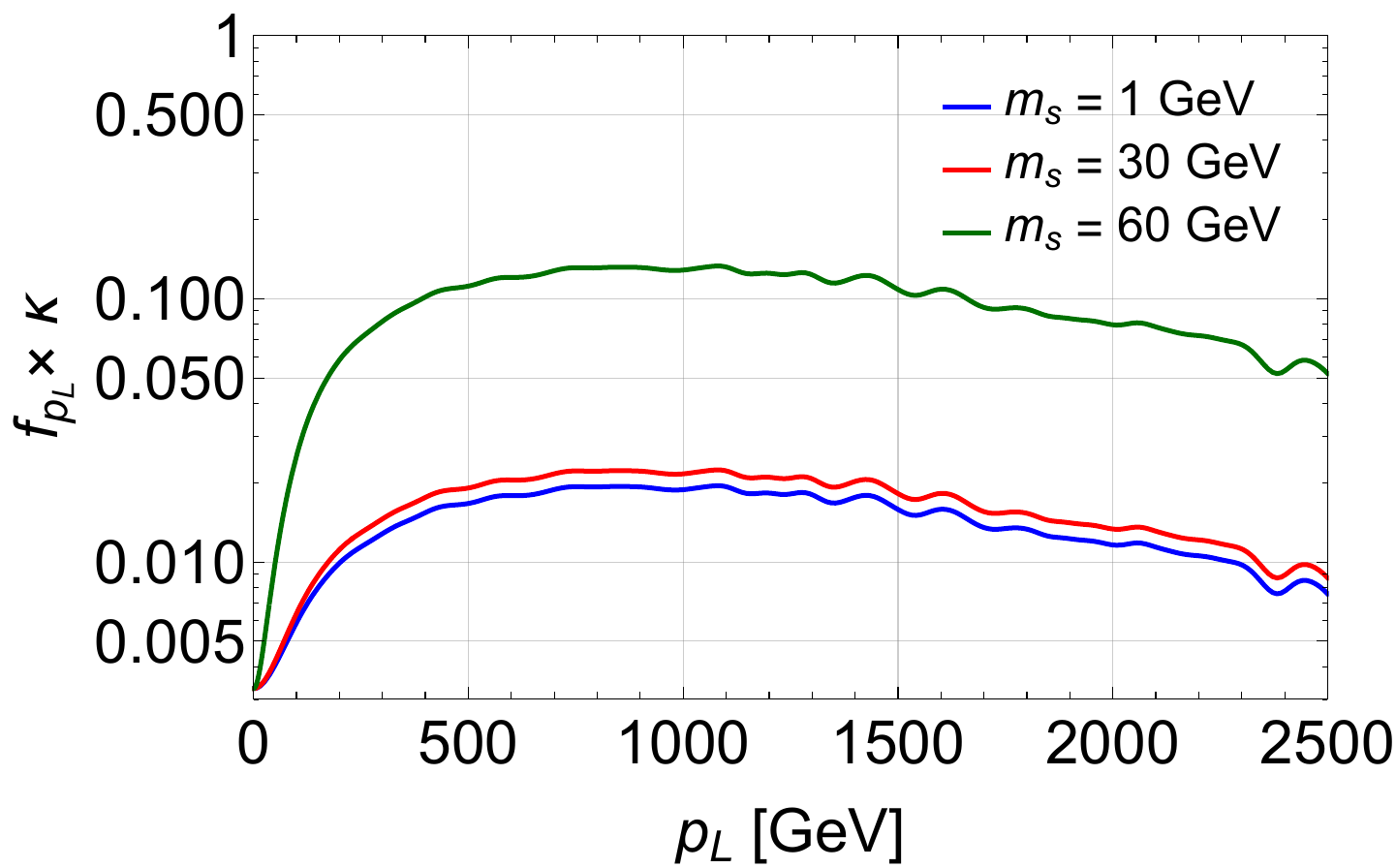}
  \caption{\textit{Left panel}: Geometric acceptance of scalars at \FASER2 obtained using $p_{L}$ distribution of Higgs bosons, see Eq.~\eqref{eq:geom-acceptance}. \textit{Right panel}: the distribution function of Higgs bosons by the longitudinal momentum $p_L$ multiplied by the enhancement factor $\kappa$~\eqref{eq:kappa} for the masses of the scalar $m_{S} = 0$, $50$ and $60$~GeV.}
  \label{fig:kappa}
\end{figure}

Based on these considerations, we can calculate the geometric acceptance (once again assuming that all Higgs bosons fly in the direction of the beam):
\begin{equation}
  \epsilon_\geom \approx \int f_{p_{L}} \kappa (m_{S},p_L) \frac{\Omega(p_{L})}{4\pi}dp_L 
  \label{eq:geom-acceptance}
\end{equation}
Here, $\Omega$ is the solid angle of \FASER2 available for scalars:

\begin{equation}
\Omega=\begin{cases}\Omega_{\faser}, \quad
\theta_{\faser}<\theta_{\text{max}}, \\ \pi \theta^{2}_{\text{max}}(p_L), \quad \theta_{\faser}>\theta_{\text{max}}, \end{cases}
\end{equation}
with $\theta_{\text{max}}=
\arctan\left[ \frac{\beta_{S}^{'}}{\gamma_{h}\sqrt{\beta_{h}^{2}-\beta_{S}^{'2}}}\right]$ if $\beta_{h}>\beta_{S}$ and $\theta_{\text{max}} = \pi$ otherwise. Finally, the function $\kappa = |d\Omega'/d\Omega|$, where $\Omega$ is the solid angle in the lab frame corresponding to the solid angle $\Omega'$ in the Higgs rest frame. It defines how collimated is the beam of scalars as compared to an isotropic distribution. For the details of the derivation of the explicit expression of $\kappa$ see Appendix~\ref{sec:kappa}. In the case $\theta = 0$ it becomes
\begin{equation}
  \kappa (m_{S},p_L)\approx \begin{cases} \frac{2\gamma_h^2(\beta_S^{'2} + \beta_{h}^{2})}{\beta_S^{'2}}, \quad \beta_{h}>\beta_{S}^{'},\\ \frac{\gamma_h^2(\beta_S^{'} + \beta_{h})^{2}}{\beta_S^{'2}}, \quad \beta_{h} < \beta_{S}^{'}\end{cases}
  \label{eq:kappa}
\end{equation}
The resulting acceptance (see Fig.~\ref{fig:kappa}, left panel) grows with the mass since the maximal angle $\theta_S$ decreases; when the mass of the scalar is very close to $m_{h}/2$, the acceptance reaches its maximum equal to the fraction of Higgs bosons flying into the direction of the \FASER2 decay volume, $f_{h\to\faser}$. Even for the light scalars the acceptance $\epsilon_{\text{geom}} \approx 4\cdot 10^{-5}$ is an order of magnitude larger than the naive estimate~\eqref{eq:N_naive}. The reason for this is that most of the Higgs bosons have large energies, so the resulting angular distribution of scalars is peaked in the direction of small angles, see Fig.~\ref{fig:thetaS_av}.

With $p_{L}$ distribution only, obviously, $f_{h\to\faser} = 1$.
To make realistic estimates, we need to take into account the $p_T$ distribution of the Higgs bosons.
The fraction $f_{h\to\faser}$ under the assumption that $p_L$ and $p_T$ distributions of Higgs boson are independent is
\begin{equation}
  f_{h\to\faser} \approx \epsilon_\geom^{\max} = \frac{1}{2} \int\limits_0^{\infty} f_{p_{L}} dp_L 
  \int\limits_0^{p_L^h \theta_\faser} f_{p_{T}} dp_T\approx 1.1\cdot 10^{-3},
  \label{eq:fhFASER}
\end{equation}
where a factor $1/2$ comes from the fact that we do not take into account Higgs bosons that fly in the opposite direction to \FASER.
This number represents a maximally possible geometric acceptance.

\subsection{Decay of scalars}
\label{sec:decay}

The decay width and branching $S\to\text{visible}$ is determined based on the (extended) results of Ref.~\cite{Boiarska:2019jym} (see Fig.~\ref{fig:decays}).
For these masses, all major decay channels have $\geqslant 2$ charged tracks and therefore it is reasonable to assume that $\BR_{\text{visible}} = 100\%$ and that every decay is reconstructable with $100\%$ efficiency. The verification of this assumption requires detailed studies beyond the scope of this paper.

So far we have kept the decay probability at its maximum (corresponding to $l_\decay = L$). This condition would give a line in the $(m_S,\theta)$ plane. To determine the transversal shape of the sensitivity region, we need to vary $\theta$ and take into account the $\gamma$ factor of the scalar, $\gamma_S$.
The energy of a scalar is proportional to the energy ($p_L$) of the
corresponding Higgs boson:
\begin{equation}
  E_S = \frac{E_{h}}{2} (1 + \beta'_{S}\beta_{h} \cos(\theta_{S})) \approx \frac{E_h}{2} (1 + \beta'_S \beta_h),
  \label{eq:ES}
\end{equation}
where we have taken into account that the \FASER detector is almost co-aligned with the beam axis and therefore $\theta_S \approx 0$ and neglected the $p_{T}$ distribution of the Higgs boson.
The average energy of the scalar is determined by weighting the Higgs distribution $f_{p_{L}}$ with the function $\kappa$, defined in Eq.~\eqref{eq:kappa}.
In this way, only the energies of scalars flying into the \FASER2 solid angle are considered. 
The resulting $\langle E_S\rangle$ as a function of the scalar mass is shown in Fig.~\ref{fig:faser2-details} (central panel). 
One can see that the $\gamma$ factor ranges from $\mathcal{O}(100)$ for small masses down to $\mathcal{O}(10)$ for $m_S \approx m_h/2$. 

Let us now improve the estimate~\eqref{eq:Pdecaymax} of the maximally possible value of the decay probability $P_{\decay}^{(\text{max})}$. The value~\eqref{eq:Pdecaymax} is obtained using the average energy $\langle E_{S}\rangle$. Taking into account the continuous scalar spectra leads to a decrease of $P_{\decay}^{(\text{max})}$. The averaging over the spectrum can be done using the function $\kappa f_{p_L}$ (shown in the right panel of Fig.~\ref{fig:kappa}):
\begin{equation}
   \langle P_{\decay}^{(\text{max})}\rangle \approx \int \kappa(m_{S},p_{L}) \cdot f_{p_{L}}\cdot P_{\decay}(m_{S},\theta^{2},E_{S})dp_{L}
\end{equation}
As is demonstrated by Fig.~\ref{fig:kappa}, $\kappa \cdot f_{p_{L}}$ have similar flat shape for wide range of momenta for all possible scalar masses. We can always adjust the appropriate $\theta^{2}$ value to maximize the probability, and independently on the mass we get
\begin{equation}
  \label{eq:2}
  \langle P_\decay^{(\text{max})} \rangle \simeq 3.2\cdot 10^{-3}
\end{equation}

\begin{figure}[t!]
    \centering
    \includegraphics[width=0.5\textwidth]{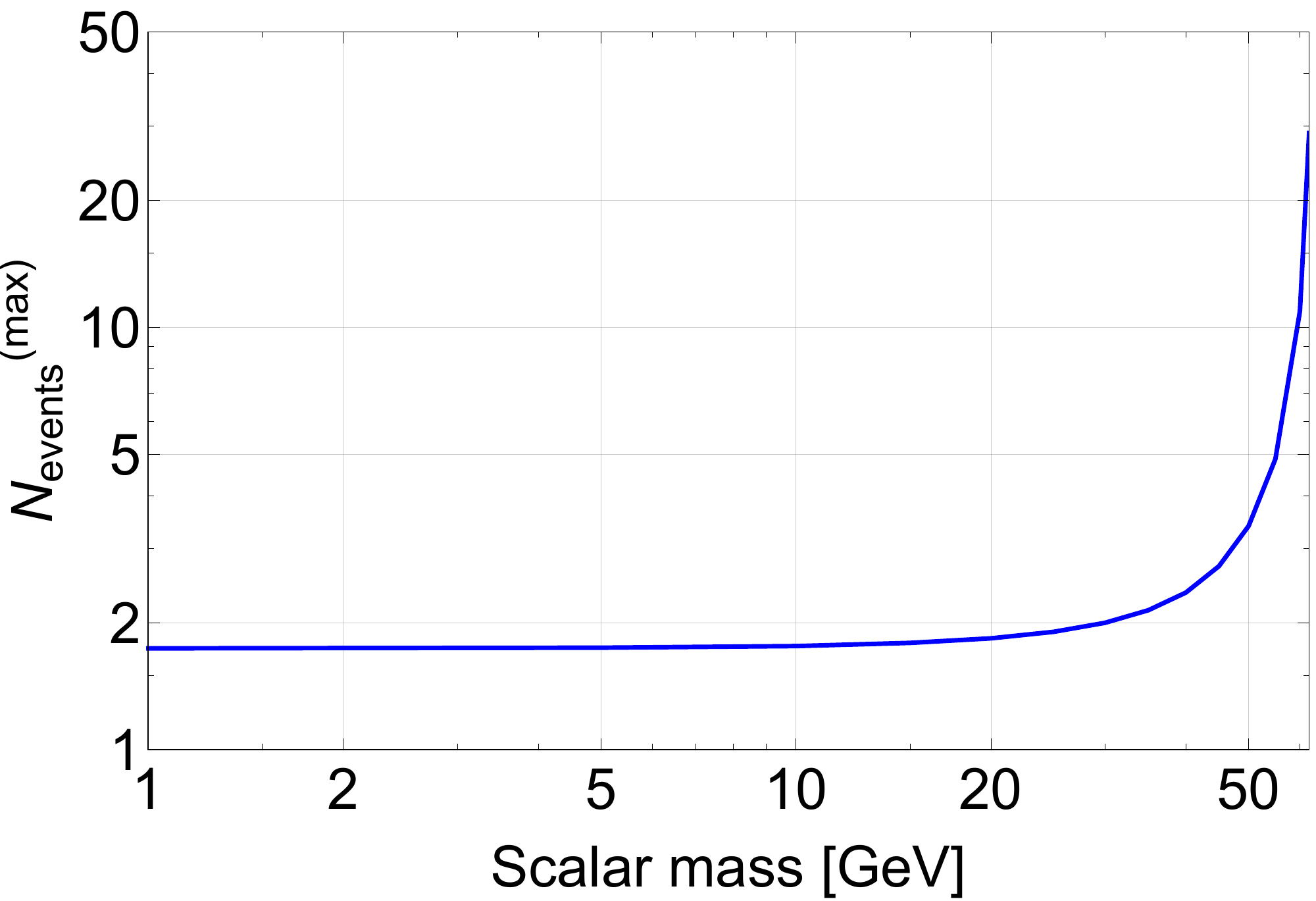}
    \caption{The analytic estimate~\eqref{eq:max-events-number} for the maximal number of scalar decays in FASER2 decay volume versus the scalar mass. See text for details.}
    \label{fig:maximal-events-number}
\end{figure}

Substituting this value for the decay probability, as well as the number of Higgs bosons produced by the fiducial branching ratio~\eqref{eq:Brinv}, $\epsilon_\geom$ (Fig.~\ref{fig:kappa}, left panel) into Eq.~\eqref{eq:1}, one can compute the improved analytic estimate for the maximal number of decay events inside the \FASER2 detector:
\begin{equation}
    N_{\text{events}}^{(\text{max})} =N_{h}\cdot \BRfid(h\to SS) \cdot  \epsilon_{\geom}\cdot\langle P_\decay^{(\text{max})}\rangle 
    \label{eq:max-events-number}
\end{equation}
It is shown in Fig.~\ref{fig:maximal-events-number}. The behavior of $N_{\text{events}}^{\text{(max)}}$ with the scalar mass is completely determined by $\epsilon_{\geom}$. Namely, the masses $m_{S}\lesssim 30\text{ GeV}$ it is a constant of the order of $\mathcal{O}(1)$, while for larger masses increases due to the behavior of the geometric
acceptance.

However, these estimates warrant a more detailed sensitivity study using the realistic distribution of Higgs bosons.

\section{Results}
\label{sec:results}

We simulated Higgs boson production at the LHC using
MadGraph5\_aMC\@NLO~\cite{Alwall:2014hca} and
following~\cite{Hirschi:2015iia}, see Appendix~\ref{sec:higgs} for details. Using the $p_L$ and $p_T$ distributions of the Higgs bosons, we derived the energy distribution of scalars $f_{E_{S}} = \frac{1}{N_{S}}\frac{dN_{S}}{dE_{S}}$ and computed the geometric acceptance $\epsilon_{\text{geom}}$, see Appendix~\ref{sec:scalar-distr}.

\begin{figure}[!t]
  \centering
    \resizebox{\textwidth}{!}{\includegraphics[height=4cm]{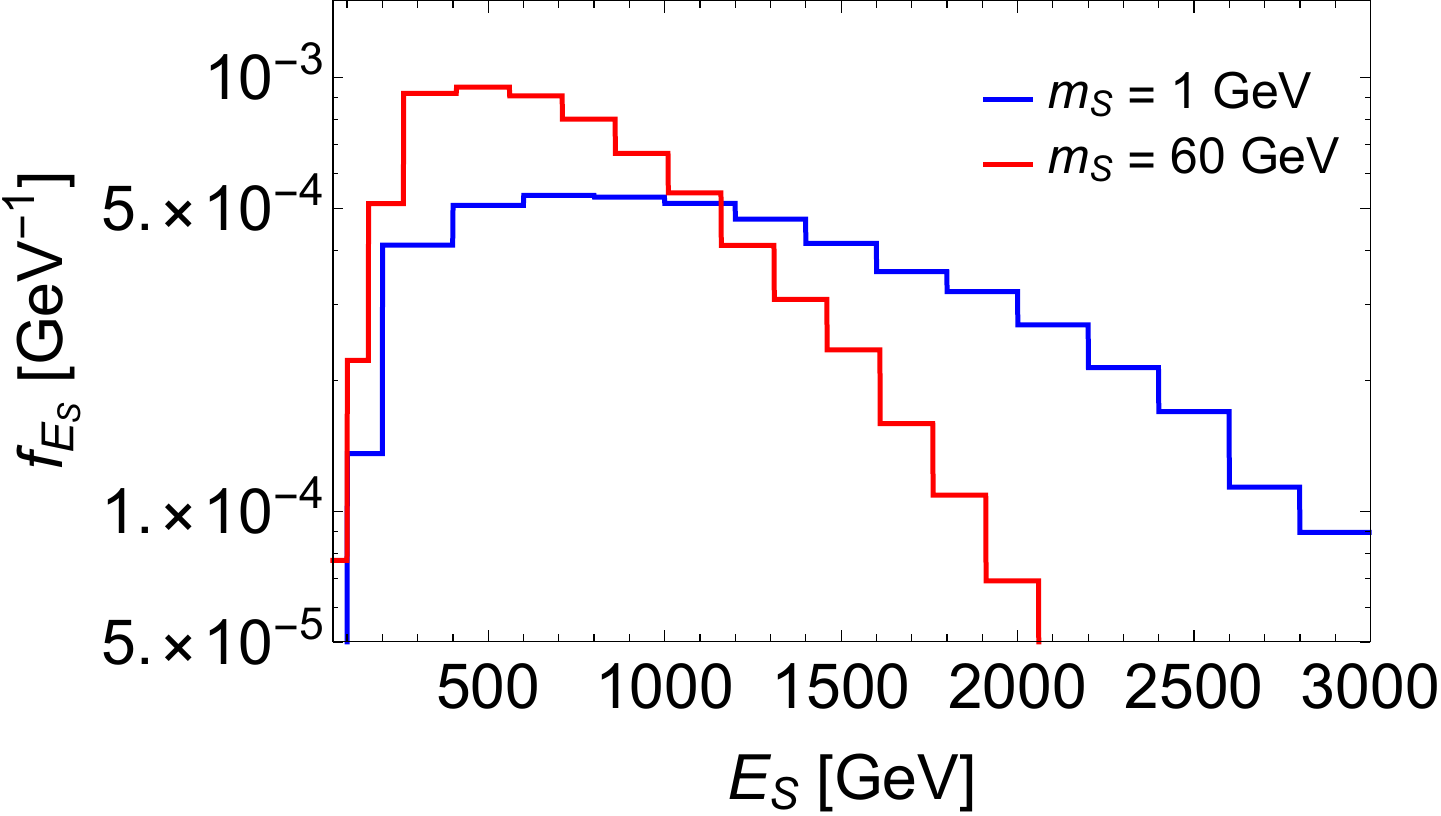}~
    \includegraphics[height=4cm]{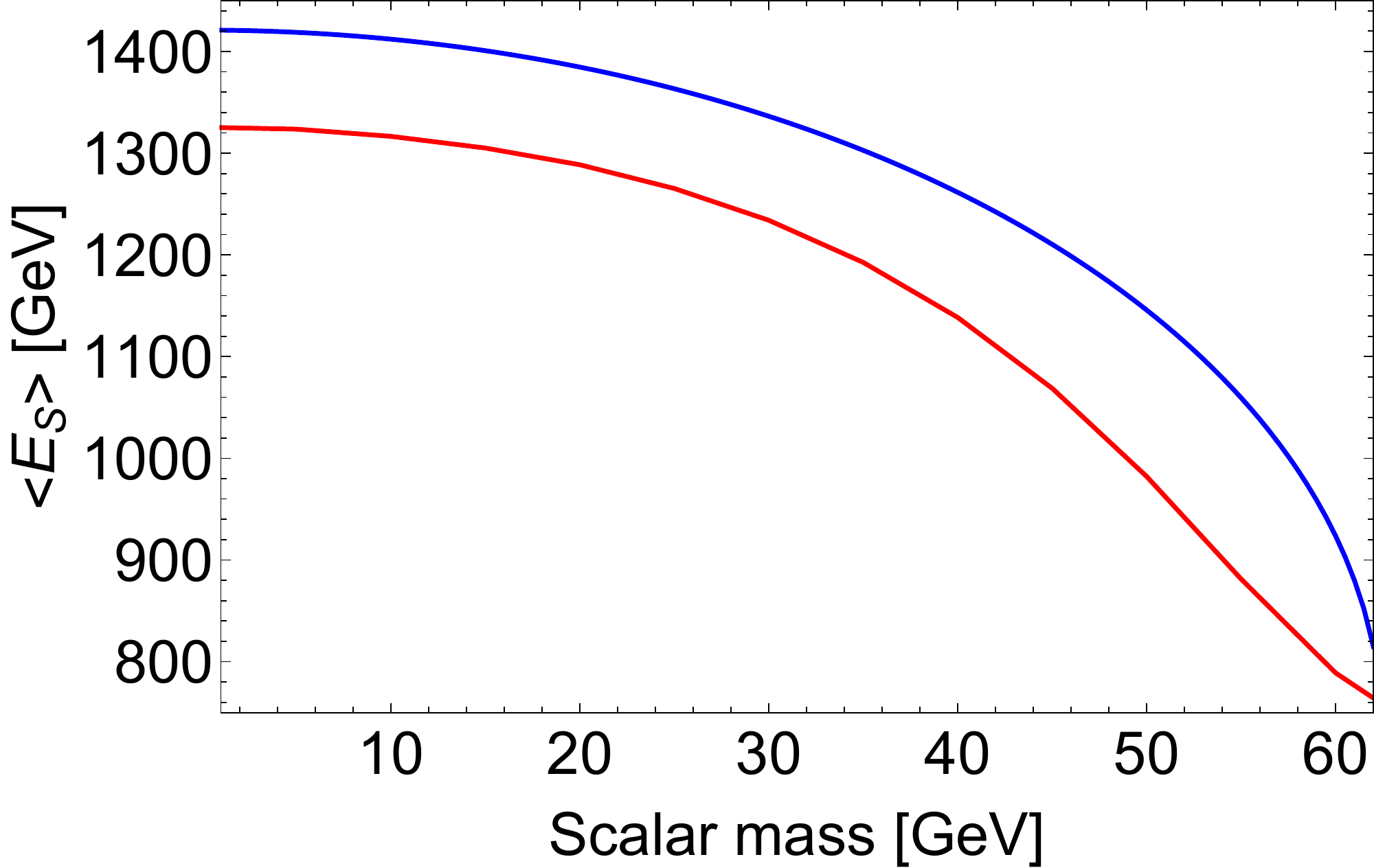}~
    \includegraphics[height=4cm]{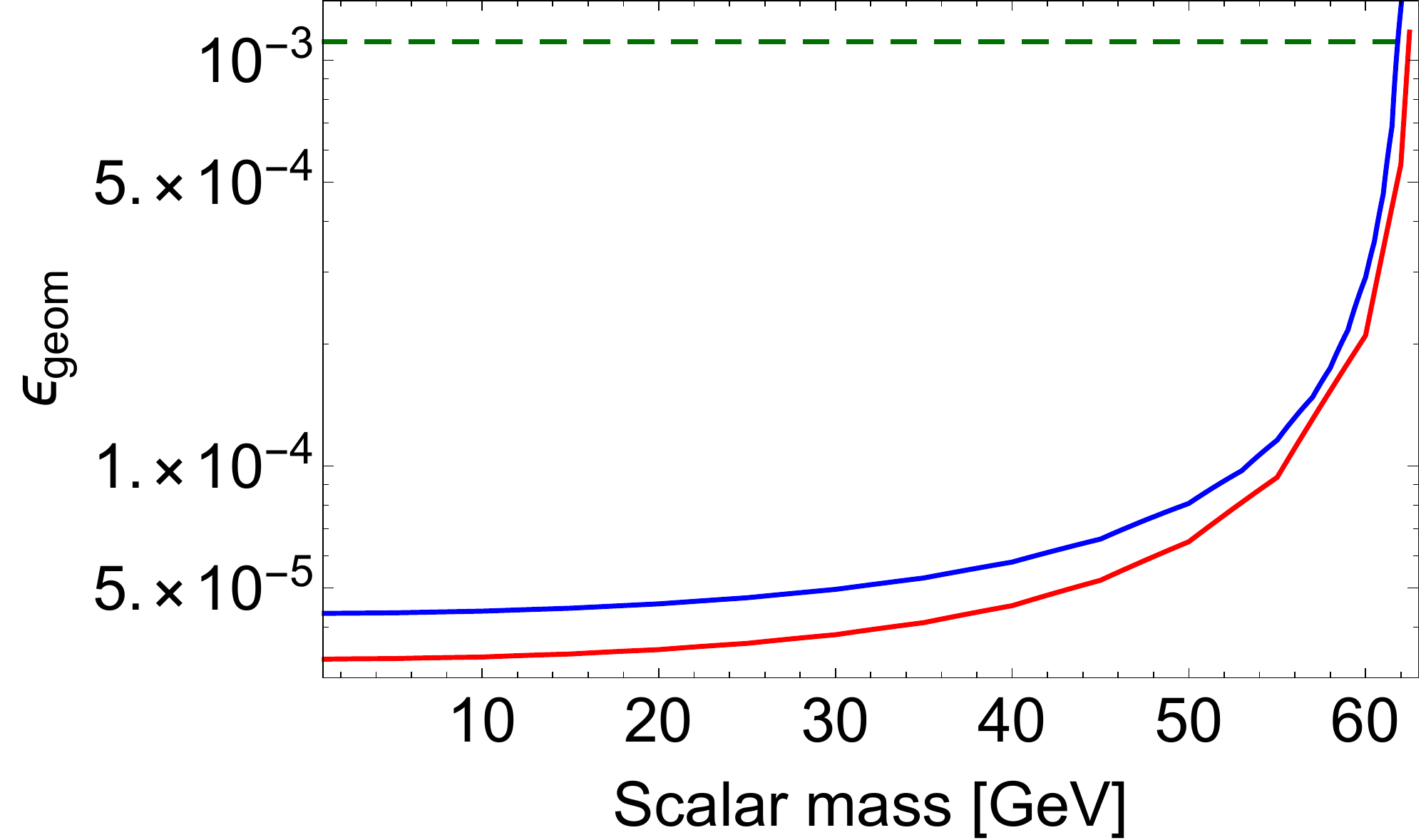}}
  \caption{Properties of dark scalars flying into the \FASER2 decay volume. \textit{Left panel}: energy spectrum of scalars $f_{E_{S}} = \frac{1}{N_{S}}\frac{dN_{S}}{dE_{S}}$ for different masses.
    \textit{Middle panel}: The average energy of scalars.
    \textit{Right panel}: The geometric acceptance $\epsilon_\geom$ versus the scalar mass. In the middle and right panels, the blue lines denote analytic estimates obtained using the Higgs $p_{L}$ spectrum (right panel in Fig.~\protect\ref{fig:Higgs-distributions}), while the red lines show the results of more accurate estimates including the $p_{T}$ distribution of the Higgs bosons (see Appendix~\ref{sec:scalar-distr}).}
  \label{fig:faser2-details}
\end{figure}

\begin{figure}[!t]
  \centering
  \includegraphics[width=0.7\textwidth]{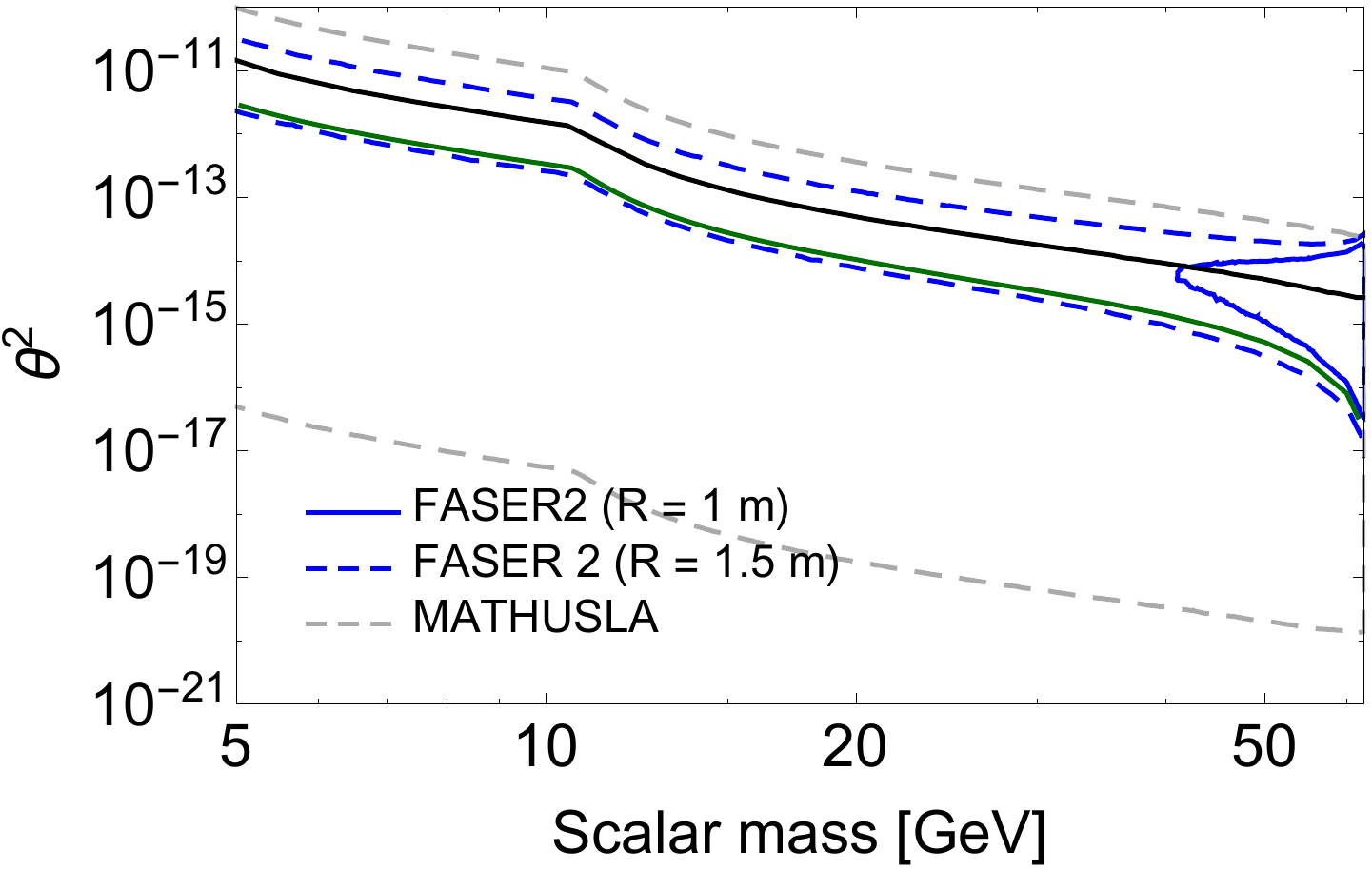}
  \caption{Sensitivity of the \FASER2 to scalars produced in decays of Higgs bosons. Blue solid line encloses the region where one expects to observe at least 2.3 events, given the current configuration of the experiment (the radius of the decay vessel $R = \unit[1]{m}$). A modest increase of the geometric acceptance (by changing the radius to $R = \unit[1.5]{m}$) allows probing an order-of-magnitude-wide stripe for all masses (between blue dashed lines).  The black solid line shows parameters for which $l_\decay = L$ (used for our analytic estimates). Gray dashed line shows upper and lower regions of the MATHUSLA200 experiment where similar production from the Higgs bosons is possible (partially based on \protect\cite{Beacham:2019nyx}). The green line is an analytic estimate, see text for details. Sensitivity estimates assume the $100\%$ efficiency of the reconstruction of decay products but take into account geometric acceptance. The branching ratio $\BR(h\to SS)$ is taken at the level of $5\%$.}
  \label{fig:faser2-sensitivity}
\end{figure}

The resulting energy distribution of scalars of particular masses traveling into the solid angle of \FASER2 is shown in Fig.~\ref{fig:faser2-details} (left panel). In the same figure (middle and right panels) we compare the geometric acceptance and average energy for scalars obtained in simulations with the analytic prediction from Fig.~\ref{fig:kappa}. The simulation results lie slightly below the analytic estimate due to the $p_{T}$ distribution of Higgs bosons. The smallness of the discrepancy is related to the smallness of the ratio $\langle p_{T}\rangle/\langle p_{L}\rangle$ for the Higgs bosons.
Next, we compute the number of scalars traveling
through the \FASER2 fiducial volume and estimate the number of decay events, using Eq.~\eqref{eq:1} with the decay probability $P_{\text{decay}}$ averaged over the energies of scalars flying in the direction of the experiment. The resulting sensitivity region
is shown in Fig.~\ref{fig:faser2-sensitivity}.
\emph{We assume background free experiment and therefore determine the sensitivity as a region that includes at least 2.3 events.}
With the current configuration of \FASER2, one can expect to see any events only in the region around $50-60$~GeV.
The green line follows from the analytic estimate~\eqref{eq:1} in which the geometric acceptance and average energy from Fig.~\ref{fig:faser2-details}, whereas the blue contours are based on the more accurate estimate using the scalar energy spectrum (see Appendix~\ref{sec:scalar-distr}). A slight difference between these estimates is caused by the difference between the value of $P_{\decay}(\langle l_\decay\rangle))$ and $\langle P_{\decay}\rangle$ where in the former case $l_\decay$ is evaluated for $\langle E_S\rangle$ and in the latter case one averages $P_\decay$ over the energy distribution. 

Our results lead to an important conclusion regarding a configuration of the \mbox{\FASER2}.
Fig.~\ref{fig:maximal-events-number} shows that the \FASER2 in its current configuration (as shown in Table~\ref{tab:FASER-parameters}) will not detect any events for $m_{N} \lesssim 40\text{ GeV}$ (region to the left of the blue solid line).
However, a modest (factor of 2) increase in the geometrical acceptance would allow probing \emph{the whole mass range} few~GeV${}\lesssim m_{S} \lesssim m_h/2$, as demonstrated by the blue dashed line in Fig.~\ref{fig:faser2-sensitivity}.
This increase can be achieved for example by increasing the radius of the \FASER2 from $\unit[1]{meter}$ to $\unit[1.5]{meters}$, which is allowed by the size of the TI12 tunnel where the experiment will be located. The angular distribution of scalars is flat for relevant angles, see Fig.~\ref{fig:thetaS_av}, which provides the desired conclusion.

\section{Conclusion}
\label{sec:conclusion}
In this work, we presented the analytic estimates for the sensitivity of the \FASER2 experiment for the most general scalar portal model including renormalizable operators only.
The estimates were verified by MadGraph simulations, showing a very good agreement.
Majority of previous works on the subject~\cite{Bird:2004ts,Batell:2009jf,Bezrukov:2009yw,Clarke:2013aya,Bezrukov:2018yvd,Winkler:2018qyg,Boiarska:2019jym} considered the models of the scalar where the term $\alpha_1 S H^{\dagger} H$ was absent in the Lagrangian~\eqref{eq:L1} (assuming a $\mathbbm{Z}_2$ symmetry $S \to -S$).
In this case, two scalar couplings $\theta$ and $\alpha$ in the effective Lagrangian~\eqref{g01} become related (and should both be small to satisfy bounds from the previous experiments).

However, if cubic and quartic couplings ($\alpha_1$ and $\alpha_2$ in the Lagrangian~\eqref{eq:L1}) are independent and both non-zero, the resulting triple coupling between Higgs and two scalars can be quite sizeable.
Indeed, the main experimental bound on its value is the branching fraction of the invisible Higgs decay (assuming it is saturated by the $h \to SS$ process).
The current bound on the invisible branching ratio $\BRinv < 0.19$ (at 95\%CL, \cite{Sirunyan:2018mtv}).
Future runs of the LHC are expected to probe this branching at the level $0.1$ or slightly below.

As a result, for the experimentally admissible values of the parameter $\alpha$, the production of scalars at the LHC from the decays of the Higgs boson ($h \to SS$) dominates significantly over all other production channels.
This makes the production and decay of a scalar controlled by independent coupling constants.
This independence qualitatively changes the behavior of the sensitivity curves of the LHC-based intensity frontier experiments (MATHUSLA, FASER, CODEX-b).
Indeed, normally the sensitivity of the intensity frontier experiments has a lower bound, defined by the minimal number of events in the detector, depends both on the production and decay, and an upper bound, defined by the requirement that new particles should not decay before reaching the detector
(the lifetime gets smaller with mass). Their intersection often defines the maximal mass of scalar that can be probed~\cite{Bondarenko:2019yob}. In our case, the maximal mass is determined solely by the kinematics ($m_S \le m_h/2$). 
However, as the geometrical acceptance drops with the decrease of the scalar's mass (see left panel of Fig.~\ref{fig:kappa}) while the number of produced scalars is mass-independent, for a given geometry there can be a  \emph{minimal mass} that can be probed (\textit{c.f.} the blue solid line in Fig.~\ref{fig:faser2-sensitivity}).

For our analysis, we assumed that the invisible Higgs decay has a significant contribution from $h\to SS$ and, as an example, adopted a fiducial branching fraction $\BR(h\to SS)$ at the level of 5\%.
We show that in this case, even if the HL-LHC does not discover invisible Higgs decay, the \FASER2 experiment is capable of discovering dark scalars with masses of $40\text{ GeV}\lesssim m_S \lesssim m_h/2$. 
Moreover, \emph{if its geometric acceptance is increased by a factor $\sim 2$}, \FASER2 will have sensitivity for all scalar masses from $m_h/2$ down to a $\text{few GeV}$ and even lower, where the production from $B$ mesons starts to contribute.
This can be achieved, for example, by scaling the radius of the detector from 1~meter to 1.5~meters.

Another possibility would be to put the detector \emph{closer} to the interaction point, in which case the number of particles, counterintuitively, increases as $L^3$ ($L^2$ dependence comes from the increase of the solid angle $\Omega_{\text{FASER}}$ and an extra factor comes from the $L$-dependence on the maximal decay probability, Eq.~\eqref{eq:Pdecaymax}).
The latter effect is due to the independence of the decay probability on the coupling $\alpha$ controlling production and is specific for the model in question.
As suggested \emph{e.g.}
in the original FASER paper~\cite{Feng:2017uoz}, another possibility would be to put the detector at 150~meters behind the TAN neutral particle absorber~\cite{Adriani:2015iwv}.
Such a position, however, would suffer from a high background and therefore our estimates (performed under the background-free assumption) will not be valid.
Another option suggested in~\cite{Feng:2017uoz} does not increase acceptance. Indeed, it was proposed to use a hollow cylinder around the beam axis, with an inner angle around 1~mrad (the size being dictated by the position of TAS quadrupole magnets shield) and the outer size of about 2~mrad.
Such a detector would have a factor of a few lower geometric acceptance.
Of course, such a detector would be too complicated and cumbersome, so its realistic version, occupying only a small sector in the azimuthal angle $\Delta \phi$, would have its geometric acceptance further reduced by $\Delta \phi/2\pi$.

\subsubsection*{Acknowledgements.}

We thank J.~Boyd, O.~Mattelaer, and S.~Trojanowski for fruitful discussions. 
This project has received funding from the European Research Council (ERC) under the European Union's
Horizon 2020 research and innovation program (GA 694896), from the
Netherlands Science Foundation (NWO/OCW). O.R.\ also acknowledges support from the Carlsberg Foundation.

\appendix

\section{Higgs boson distribution}
\label{sec:higgs}

\begin{figure}[h!]
  \centering
  \includegraphics[width=0.5\textwidth]{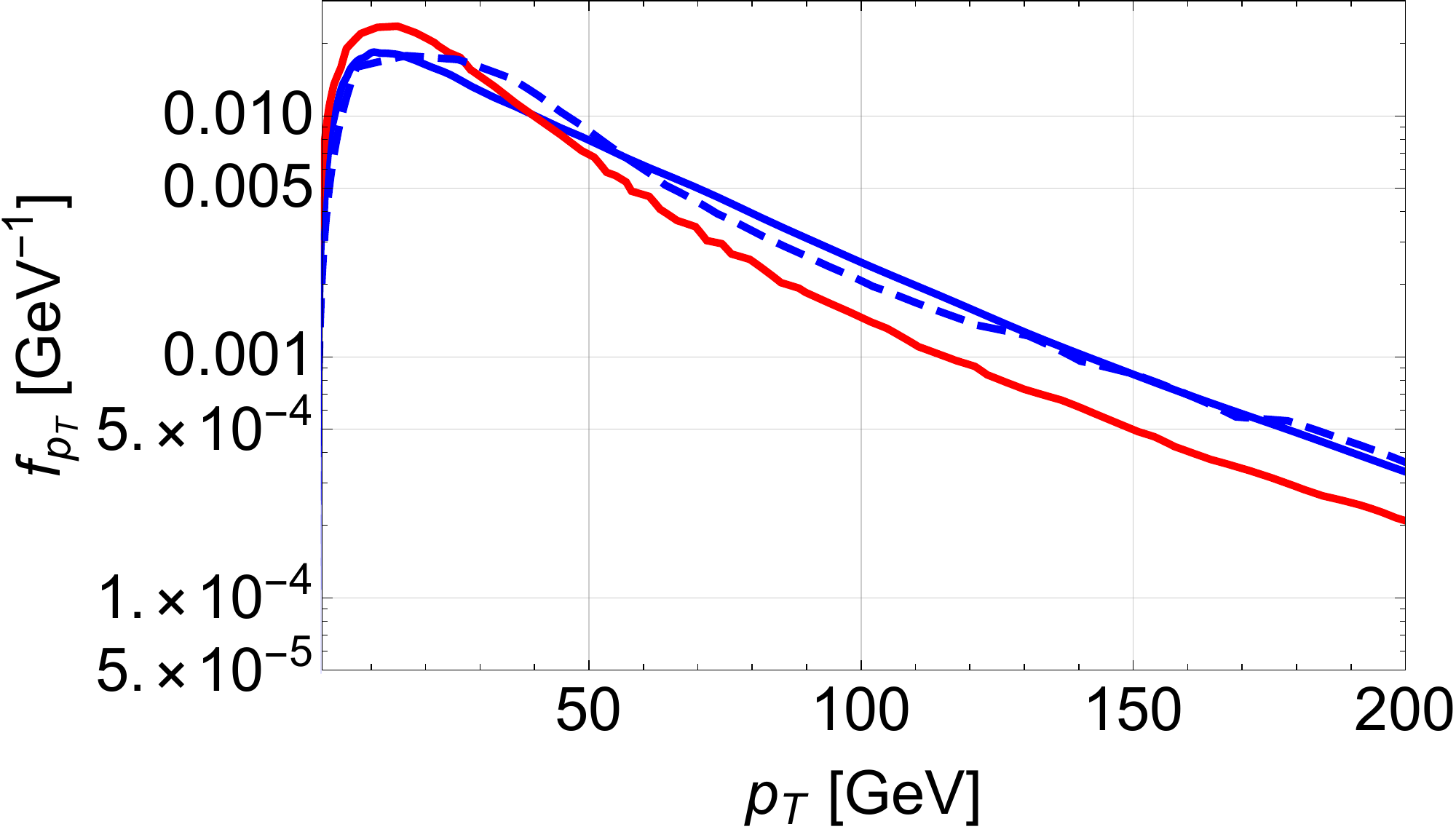}~\includegraphics[width=0.5\textwidth]{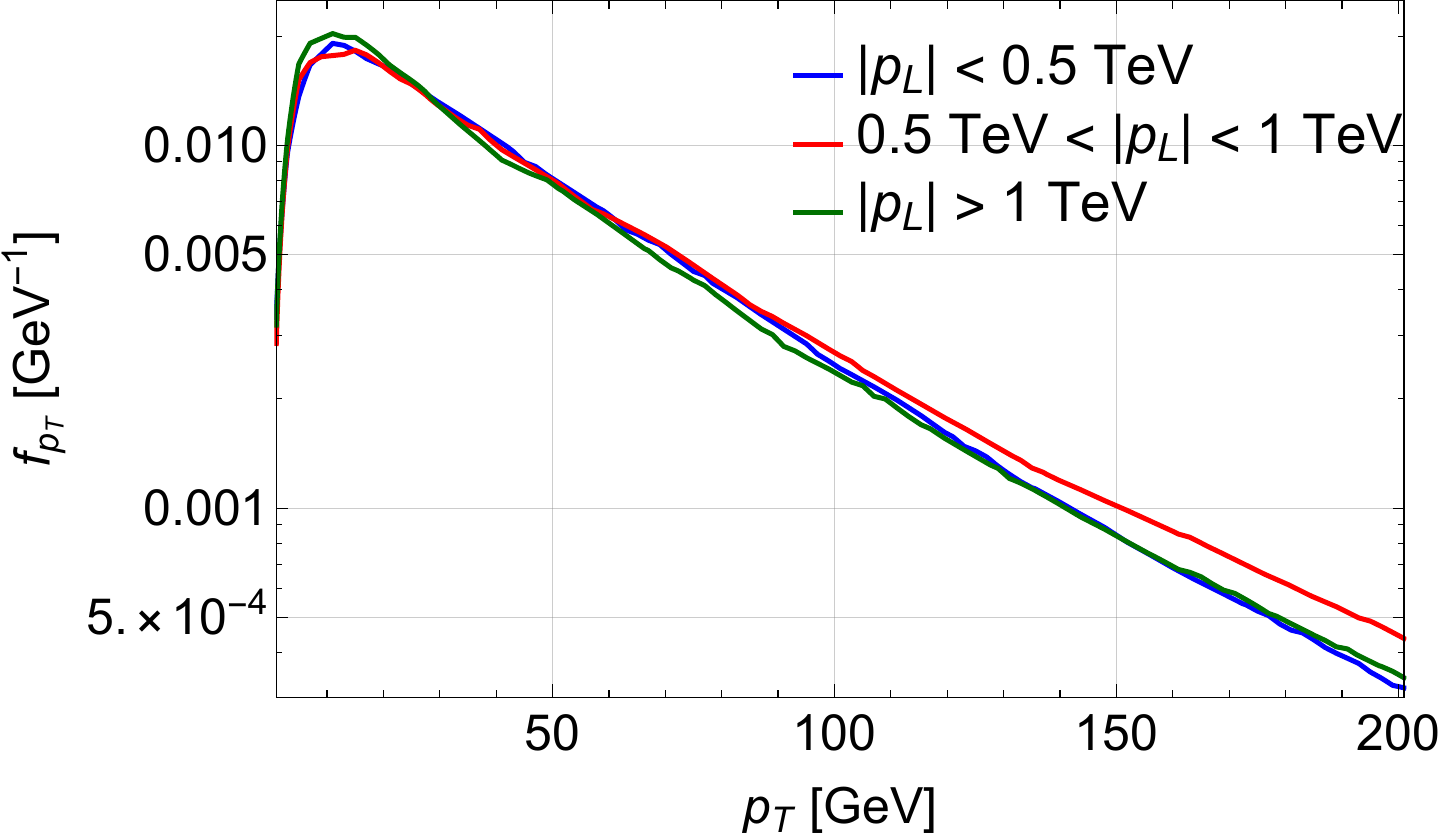}
  \caption{\textit{Left panel}: a comparison of $p_{T}$ spectra of Higgs
    bosons obtained in our simulations (solid blue line) with the spectra
    from~\cite{Hirschi:2015iia} (dashed blue line) and~\cite{Bagnaschi:2015bop} (red
    line). See text for details. \textit{Right panel}: the $p_{T}$
    distribution of Higgs bosons for different domains of $|p_{L}|$.}
  \label{fig:pt-comparison}
\end{figure}

For our estimate we used a number of Higgs bosons for HL LHC $N_h = 1.7\cdot 10^{8}$. To find Higgs bosons momentum distribution, we simulated Higgs boson production at the LHC using
MadGraph5\_aMC\@NLO~\cite{Alwall:2014hca} and
following~\cite{Hirschi:2015iia}.  Using the generated events, we find that
the $p_{T}$ distribution depends only weakly on $p_{L}$, see
Fig.~\ref{fig:pt-comparison}.  Therefore, the correlations between $p_{T}$ and
$p_{L}$ distributions can be neglected, and the double distribution of Higgs
bosons in $p_{T},p_{L}$ can be approximated by the product of $p_{T}$ and
$p_{L}$ single distributions.

We validated our simulation by comparing the $p_{T}$ spectrum of the Higgs
bosons with the theoretical spectra from~\cite{Hirschi:2015iia}
and~\cite{Bagnaschi:2015bop}, in which the spectrum was obtained using POWHEG,
see Fig.~\ref{fig:pt-comparison}.
Our results agree well with~\cite{Hirschi:2015iia}, while there is a
discrepancy with~\cite{Bagnaschi:2015bop} in the domain of high
$p_{T}$. However, the discrepancy is not significant; in particular, the
amounts of Higgs bosons flying in the direction of FASER 2 experiment
calculated using our distribution and the distribution
from~\cite{Bagnaschi:2015bop} differs by no more than 30\%.

For each simulated event we calculated $\kappa(\theta_{h}, \gamma_{h})$ and
the energy $E_{S}(\theta_{h},\gamma_{h})$ of a scalar traveling into the solid
angle of \FASER2.  The $\langle \kappa\rangle$ is then obtained as the
arithmetic mean, while the energy distribution is obtained as the weighted
distribution, where the energy $E_{S}(\theta_{h},\gamma_{h})$ has the
corresponding weight $\kappa(\theta_{h},\gamma_{h})$.

\section{Distributions}
\subsection{Kinematics in laboratory frame}
\label{sec:kappa-calculation}
Consider the relation between the laboratory frame angle $\theta_{S}$ and the rest frame angle $\theta^{'}_{S}$:
\begin{equation}
    \tan(\theta_{S}) = \frac{1}{\gamma_{h}}\frac{\beta^{'}_{S}\sin(\theta^{'}_{S})}{\beta^{'}_{S}\cos(\theta_{S}^{'})+\beta_{h}}
    \label{eq:kinematic-angles}
\end{equation}
Let us introduce two functions
\begin{equation}
    f_{\pm}(\theta_{S}) = -\frac{\beta_{h}\gamma_{h}^{2}\tan^{2}(\theta_{S})\pm\sqrt{\beta_{S}^{'2}+(\beta_{S}^{'2}-\beta_{h}^{2})\gamma_{h}^{2}\tan^{2}(\theta_{S})}}{\beta_{S}^{'}(1+\gamma_{h}^{2}\tan^{2}(\theta_{S}))}, 
    \label{eq:angle-solution}
\end{equation}
representing the solution  of Eq.~\eqref{eq:kinematic-angles} in terms of $\cos(\theta_{S}^{'})$ for given  parameters $\beta_{h},\beta_{S}$. 
In order to express $\cos(\theta_{S}^{'})$ from Eq.~\eqref{eq:kinematic-angles}, we find first the values of $\theta_{S}$ where the functions~\eqref{eq:angle-solution} become complex. These are $\theta_{S,\text{max}} < \theta_{S}< \pi - \theta_{S,\text{max}}$, defined as
\begin{equation}
\theta_{S,\text{max}} = \arctan\left[\frac{\beta_{S}^{'}}{\gamma_{h}\sqrt{\beta_{h}^{2}-\beta_{S}^{'2}}}\right]
\end{equation}
They are always real as long as $\beta_{h}/\beta_{S}^{'}<1$. Next, we can construct the physical solution $\cos(\theta_{S}^{'})$ requiring the solutions~\eqref{eq:angle-solution} to cover all the domain of the definition of the cosine, $\cos(\theta_{S}^{'})\in [-1,1]$. For $\beta_{h}/\beta_{S}^{'}<1$ it is
\begin{multline}
   \cos(\theta_{S}^{'}) = \begin{cases}f_{-}(\theta_{S}), 0<\theta_{S}<\pi/2, \\ f_{+}(\theta_{S}), \quad \pi/2 < \theta_{S}<\pi\end{cases} = \\ = -\frac{\beta_{h}\gamma_{h}^{2}\sin^{2}(\theta_{S})-\cos(\theta_{S})\sqrt{\beta_{S}^{'2}\cos^{2}(\theta_{S})+(\beta_{S}^{'2}-\beta_{h}^{2})\gamma_{h}^{2}\sin^{2}(\theta_{S})}}{\beta_{S}^{'}\cos^{2}(\theta_{S})+\gamma_{h}^{2}\sin^{2}(\theta_{S})}
   \label{eq:physical-solution-1}
\end{multline}
For $\beta_{h}>\beta_{S}^{'}$ both the solutions $f_{\pm}$ exist in the domain $\theta_{S} < \theta_{S,\text{max}}$. 

Let us now find the function $\kappa$. By the definition, $\kappa = |d\cos(\theta_{S}^{'})/d\cos(\theta_{S})|$. In the case $\beta_{h}<\beta_{S}^{'}$ it is simply given by the derivative of~\eqref{eq:physical-solution-1}, while for the case $\beta_{h}>\beta_{S}^{'}$ it reads
\begin{equation}
    \kappa = \left|\frac{df_{+}(\theta_{S})}{d\cos(\theta_{S})}\right|+\left|\frac{df_{-}(\theta_{S})}{d\cos(\theta_{S})}\right| = \frac{dg(\theta_{S})}{d\cos(\theta_{S})},
\end{equation}
where
\begin{equation}
    g(\theta_{S}) = \left|2\cos(\theta_{S})\frac{\sqrt{\beta_{S}^{'2}\cos^{2}(\theta_{S})+(\beta_{S}^{'2}-\beta_{h}^{2})\gamma_{h}^{2}\sin^{2}(\theta_{S})}}{\beta_{S}^{'}(\cos^{2}(\theta_{S})+\gamma_{h}^{2}\sin^{2}(\theta_{S}))}\right|
\end{equation}
In particular, in the domain $\theta_{S}\ll \theta_{S,\text{max}}$ for $\beta_{h}>\beta_{S}^{'}$ we have
\begin{equation}
    g(\theta_{S})\approx 2-\frac{\theta_{S}^{2}(\beta_{h}^{2}+\beta_{S}^{'2})}{\beta_{S}^{'2}} \Rightarrow \kappa \approx \frac{2\gamma_{h}^{2}(\beta_{S}^{'2}+\beta_{h}^{2})}{\beta_{S}^{'2}}
\end{equation}

\subsection{Distribution of scalars over energies and polar angles}
\label{sec:scalar-distr}
The double differential distribution $f_{E_{S},\theta_{S}}$ of scalars
produced in the decay $h\to SS$ has been calculated in the following way.
Consider a differential
branching ratio for a Higgs bosons flying in the direction $\theta_{h},\phi_{h}$:
\begin{equation}
\label{eq:BrSS}
  d\text{Br}(h\to SS) = \frac{1}{2}\frac{1}{(2\pi)^{2}}\frac{|\mathcal{M}|^{2}}{2\Gamma_{h,\text{rest}}m_{h}}\frac{d^{3}\bm{p}_{S_{1}}}{2E_{S_{1}}}\int \frac{d^{3}\bm{p}_{S_{2}}}{2E_{S_{2}}}\delta^{4}(p_{h}-p_{S_{1}}-p_{S_{2}}),
\end{equation}
where $\mathcal{M}$ is the invariant matrix element of the process
$h\to SS$ (independent on momenta for $1\to 2$ process), $\bm{p}_{S_{1,2}}$ are momenta of two produced scalars.

\begin{figure}
    \centering
    \includegraphics[width=0.5\textwidth]{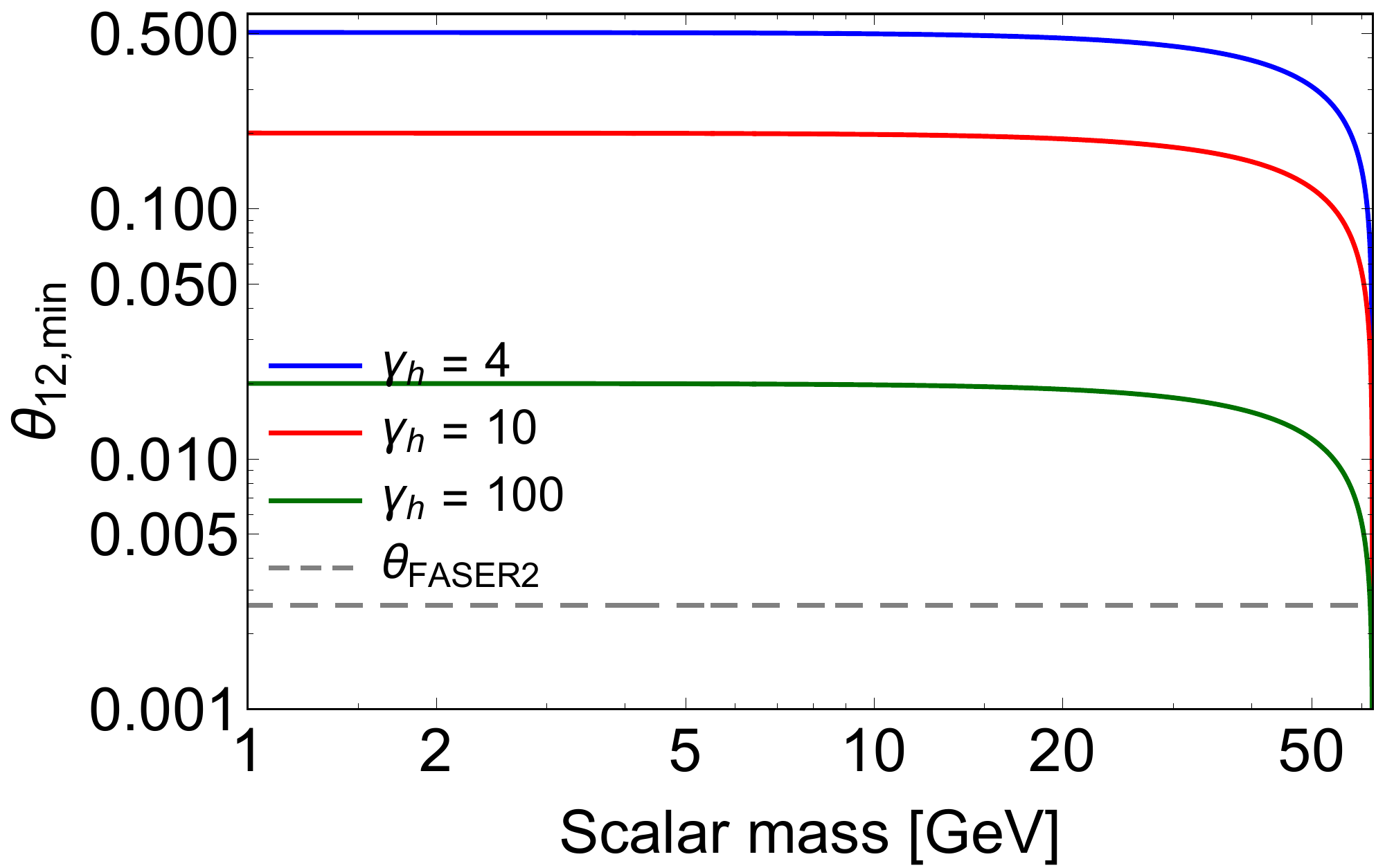}
    \caption{The minimal angle~\eqref{eq:minimal-angle} between two scalars produced in the decay $H\to SS$ versus the scalar mass $m_{S}$ for particular values of the $\gamma$ factor of the Higgs boson.}
    \label{fig:minimal-angle}
\end{figure}

Two scalars are indistinguishable (extra factor $1/2$ in Eq.~\eqref{eq:BrSS}) and after phase space integration we would lose the information about the relative distribution of the two scalars.
In particular we cannot trace whether one or both  scalars simultaneously could enter  the \FASER2 decay volume which could lead to underestimate of the number of events by as much as a factor of 2.
However, because of the small angular size of the \FASER2 experiment, the fraction of events with two $S$s flying into the detector's fiducial volume is negligibly small. Indeed, the minimal angle $\theta_{12,\text{min}}$ between two scalars produced in the decay $h\to SS$ is given by
\begin{equation}
\sin(\theta_{12,\text{min}}) = \frac{2m_{h}^{2}\beta_{h}\sqrt{\gamma_{h}^{2}-1}}{m_{h}^{2}\gamma^{2}_{h}-4m^{2}_{S}}
\label{eq:minimal-angle}
\end{equation}
It is \textit{larger} than $\theta_{\faser} \approx 2.6\cdot 10^{-3}$ for all values of $\gamma_{h}$ reachable at the LHC for $m_{S} \lesssim 62\text{ GeV}$, see Fig.~\ref{fig:minimal-angle}.
After the integration over $\mathbf p_{S_{2}}$, replacing $S_{1}\to S$ we get
  \begin{equation}
 d\text{Br}(h\to SS) = \frac{d^{3}\mathbf p_{S}}{8(2\pi)^{2}}\frac{|\mathcal{M}|^{2}}{4\Gamma_{h,\text{rest}}m_{h}E_{S}}\delta(m_{h}^{2}-2E_{S}E_{h}+2|\bm{p}_{S}||\bm{p}_{h}|\cos(\alpha)),
\end{equation}
where
\begin{equation}
  \cos(\alpha) =\cos(\phi_{h})\sin(\theta_{h})\sin(\theta_{S})+\cos(\theta_{h})\cos(\theta_{S})
\end{equation}
is the angle between the Higgs boson and the scalar. Rewriting the scalar phase space volume as
$d^{3}\bm{p}_{S} = \sin(\theta_{S})d\theta_{S}
E_{S}\sqrt{E_{S}^{2}-m_{S}^{2}}dE_{S}d\phi_{S}$, for the distribution in the energy and polar angle is given by
\begin{multline}
  f_{\theta_{S},E_{S}} = \frac{1}{\BR_{h\to SS}}\frac{d\BR (h\to SS)}{d\theta_{S}dE_{S}} = \\ = 2\pi\frac{\sin(\theta_{S})E_{S}\sqrt{E_{S}^{2}-m_{S}^{2}}}{\text{Br}(h\to SS)} \int \frac{d\phi_{h}}{2\pi}dE_{h}d\theta_{h}f_{\theta_{h},E_{h}}\frac{d^{3}\text{Br}(h\to SS)}{d^{3}\bm{p}_{S}} = \\ = \frac{m_{h}\sqrt{E_{S}^{2}-m_{S}^{2}}}{|\bm{p}_{S,\text{rest}}|}\sin(\theta_{S})I[\theta_{S},E_{S}],
  \label{eq:distribution function}
\end{multline}
where $f_{\theta_{h},E_{h}}$ is the double differential distribution of the Higgs bosons obtained in simulations, and
\begin{equation}
  I[\theta_{S},E_{S}] = \frac{1}{2\pi}\int d\phi_{h}d\theta_{h}dE_{h}f_{\theta_{h},E_{h}}\delta(m_{h}^{2}-2E_{S}E_{h} + 2|\bm{p}_{S}||\bm{p}_{h}|\cos(\alpha))
\end{equation}
Having the distribution function~\eqref{eq:distribution function}, the number of events may be determined as
\begin{equation}
N_{\det} = N_{S}\cdot \BR(h\to SS) \cdot \int d\theta_{S}dE_{S} f_{\theta_{S},E_{S}}P_{\decay}(E_{S})
\label{eq:decay-events-accurate}
\end{equation}

\bibliographystyle{JHEP}
\bibliography{ship.bib}

\end{document}